\documentclass[12pt,preprint]{aastex}

\usepackage{epstopdf}
\usepackage{natbib}
\usepackage{graphicx}

\newcommand{\tco}{\ifmmode {^{13}{\rm CO}} \else {$^{13}{\rm CO}$}\fi}
\newcommand{\dco}{\ifmmode {^{12}{\rm CO}} \else {$^{12}{\rm CO}$}\fi}
\newcommand{\juz}{\ifmmode {{\rm J}=1\rightarrow 0} \else
{J=1$\rightarrow$0}\fi}
\newcommand{\jdu}{\ifmmode {{\rm J}=2\rightarrow 1} \else
{J=2$\rightarrow$1}\fi}
\newcommand{\jtd}{\ifmmode {{\rm J}=3\rightarrow 2} \else
{J=3$\rightarrow$2} \fi}

\shorttitle{ALMA B-field in W51}
\shortauthors{Koch et al.}

\begin{document}


\title{Polarization Properties and Magnetic Field Structures in the High-Mass Star-Forming Region W51 Observed with ALMA}
\author{Patrick M. Koch$^1$, Ya-Wen Tang$^1$, Paul T.P. Ho$^{1,2}$, Hsi-Wei Yen$^3$, Yu-Nung Su$^1$, 
and Shigehisa Takakuwa$^{4,1}$}
\affil{$^{1}$Academia Sinica, Institute of Astronomy and Astrophysics, Taipei, Taiwan}
\affil{$^{2}$East Asian Observatory (EAO), 660 N. Aohoku Place, University Park, Hilo, Hawaii 96720, USA}
\affil{$^{3}$European Southern Observatory (ESO), Karl-Schwarzschild-Str. 2, D-85748 Garching, Germany}
\affil{$^{4}$Department of Physics and Astronomy, Graduate School of Science and Engineering, Kagoshima University, 1-21-35 Korimoto, Kagoshima, Kagoshima 890-0065, Japan}

\email{pmkoch@asiaa.sinica.edu.tw}

\begin{abstract}
We present the first ALMA dust polarization observations towards the high-mass star-forming regions W51 e2, e8, 
and W51 North in Band 6 (230 GHz) with a resolution around 0$\farcs$26 ($\sim5$mpc). 
Polarized emission in all three sources
is clearly detected and resolved. Measured relative polarization levels are between 0.1\% and 10\%. While the absolute
polarization shows complicated structures, the relative polarization displays the typical anti-correlation with Stokes $I$, 
though with a large scatter.
Inferred magnetic (B) field morphologies are organized and connected. Detailed substructures are resolved, revealing 
new features such as cometary-shaped B-field morphologies in satellite cores, symmetrically converging B-field zones, 
and possibly streamlined morphologies. The local B-field dispersion shows some anti-correlation with the relative polarization.
Moreover, lowest polarization percentages together with largest dispersions coincide with B-field convergence zones.
We put forward $\sin\omega$, where $\omega$ is the measurable angle between a local B-field orientation and local gravity,
as a measure of how effectively the B-field can oppose gravity. 
Maps of $\sin\omega$ 
for all three sources show organized
structures that suggest a locally varying role of the B-field, with some regions where gravity can largely act unaffectedly,
possibly in a network of narrow magnetic channels,
and other regions where the B-field can work maximally against gravity.
\end{abstract}

\section{Introduction}

The giant molecular cloud W51 is among the most massive star-forming regions in our Galaxy.
The W51 complex is further unique as it is located in a region with little foreground and background
contamination. \citet{ginsburg17} gives a recent observational review.  
The major regions within W51 are  W51 A, B, and C. The two most luminous high-mass protostars, W51 e2
and W51 North, are located in the W51 A region 
which has a luminosity equivalent to a star cluster of 5,000 to 10,000 M$_{\odot}$.
W51 e2 together with e8 is located along a molecular ridge at a parallax distance around 5.41~kpc
\citep{sato10}. Parallax measurements towards W51 North yield about 5.1~kpc \citep{xu09}.
W51 contains an aggregation of HII regions \citep{westerhout58, martin72, mehringer94} and masers
detected in several molecular lines (OH in W51 e2, e8, and North, \citep{etoka12}; H$_2$O in e2, e8, and North
\citep{genzel81, imai02, eisner02}; CH$_3$OH in e2, e8, and North \citep{phillips05, etoka12}; SiO and NH$_3$
in W51 North \citep{morita92, eisner02, brown91, gaume93, henkel13}.
W51 A, due to being
very bright in the millimetre wavelength range, has been extensively studied with a variety of molecular lines. 
Evidence for infall and / or accretion around W51 e2 is reported in \citet{ho96, zhang97,zhang98, young98, sollins04}.
A possible rotation with a spinning-up motion is discussed in \citet{zhang97,zhang98}.
Observations of the hydrogen recombination line H$53\alpha$ led to the interpretation of a rotational 
ionized accretion flow around the ultra-compact HII region in e2 \citep{keto08}. Later higher-resolution 
observations by \citet{shi10a,shi10b, goddi15,goddi16} reveal that e2 fragments at least into two sources, 
e2-west (a hyper-compact HII region) and e2-east (a hot molecular core), suggesting that the accretion is possibly onto these smaller-scale cores.
W51 e8, south of e2 at a projected distance around 0.3~pc, appears to be in a common larger-scale 0.5~pc envelope
together with e2 \citep{lai01, tang09b}. Infall signatures towards e8 are detected in NH$_3$ \citep{ho96,zhang97,zhang98}
and in CS \citep{zhang98}, indicating an early evolutionary stage. 
Towards W51 North, infall motions are detected in SO$_2$ \citep{sollins04} and CN \citep{zapata08}.
At a higher angular resolution of $0\farcs4$, \citet{zapata09} observed an infalling ring-like structure in SO$_2$. 
The orientation of the molecular outflow traced in SiO (5--4) (150$^{\circ}$, \citet{zapata09}) is similar to the 
orientation derived from proper motions of H$_2$O masers (105$^{\circ}$ to 140$^{\circ}$, \citet{eisner02, imai02}).
High-resolution continuum observations at $0\farcs7$ resolve W51 North in at least four smaller cores along 
an east-west direction south of the cometary shell-like HII region W51 d \citep{tang13}.
Recent ALMA observations with a resolution of 1000 au in multiple lines do not reveal unambiguous signatures of infall
in e2, e8, or North, but this is likely caused by observational limitations rather than a non-existence of infall motion \citep{ginsburg17b}. 

The focus of the present paper is to study the role of the magnetic (B) field with the 
Atacama Large Millimeter/submillimeter Array (ALMA) at a physical 
scale around 5~mpc in W51 e2, e8, and North.  To that purpose we are making use of dust polarization observations.
At the densities and scales probed with our resolution in these high-mass star-forming regions, dust grains are expected to be coupled to the B-field, aligned with their shorter axis parallel to the B-field. 
Their emission at (sub-)millimetre wavelength is, thus, polarized perpendicular to the B-field lines 
\citep{cudlip82, hildebrand84, hildebrand88, lazarian00, andersson15}.           
This dust alignment is likely made possible through radiative torques 
\citep{draine96, draine97, lazarian00, cho05, lazarian07,hoang16}.

The W51 e2/e8 ridge has already been the focus of dust polarization observations in order to map and study 
the B-field.  Berkeley-Illinois-Maryland-Association (BIMA) radio telescope array observations 
at 1.3~mm with a 
resolution $\theta\sim 3\arcsec$ show an elongated one-parsec long envelope around e2/e8 with the 
B-field mostly oriented perpendicular to the envelope's longer axis \citep{lai01}.
At the location of the e2 core there is noticeably less or no polarization detected with this $\sim 0.1$pc resolution.
Higher-resolution observations with the Submillimeter Array (SMA) clearly resolved a radial-like pinched
B-field morphology 
precisely at the location of e2
and a more stretched morphology in e8 with $\theta\sim0\farcs7$ at $870\mu$m \citep{tang09b}. Comparing gravitational force and B-field tension and noticing that both the e2 and e8 core show 
signatures of infall, upper limits for their B-field strengths are estimated to be $<19$~mG (e2) and 
$<8$~mG (e8), respectively \citep{tang09b}.  The large-scale B-field in the plane of the sky, pervading the 0.5-pc envelope,  is estimated to be $\sim1$~mG from the Chandreasekhar-Fermi \citep{chandra53} method in \citet{lai01}. 
Subsequently, W51 e2 served as a testbed for a newly developed technique -- the polarization-intensity gradient
method -- to measure local magnetic field strengths \citep{koch12a, koch12b}. A clear increase in field 
strengths is measured from $\sim1$~mG in the core's peripheral zones to a central value around 20~mG. 
\citet{etoka12} quote a B-field strength of 2-7~mG from OH masers in e2 which is similar to field strengths
measured in other compact HII regions detected through Zeeman observations of OH masers 
(a few mG up to $\sim20$~mG, \citet{fish07}). 
An analysis of the local magnetic field-to-gravity force ratio shows a clear drop towards the center of e2, indicating
that the B-field is largely overwhelmed by gravity in the central region while the field can still provide 
resistance (force ratio larger than unity) in the northwestern area \citep{koch12a}. 
An identical picture is seen on larger scales, both in between W51 e2/e8 and W51 North, and in between
W51 e2 and e8 with a larger field resistance against gravity in between the cores and gravity dominating 
at the locations of the cores \citep{koch12b}.

W51 North was the target of a systematic study to probe the change in B-field morphologies \citep{tang13}
from a 3-pc 
envelope surface layer (JCMT/SCUPOL at $850\mu$m, \citet{chrys02,matthews09}) 
to the pc-scale molecular cloud 
probed with the CSO/Hertz at $350\mu$m \citep{dotson10} down
to the core envelope and core resolved with the SMA at the 0.1-pc scale \citep{tang13}. 
The systematic change from a close-to-uniform larger-scale B-field, to a symmetric from north and south
channelling field morphology, down to a pulled-in hourglass-like B-field is also reflected in a tightening 
correlation between emission gradients and field orientations \citep{tang13} which is explained 
by gravity more and more dominating over the B-field \citep{koch13}. This finding was later confirmed 
in a large 50-source sample with the SMA and the CSO data in \citet{koch14}. 
In W51 North, field-to-gravity force ratios are small (around 0.5) on average, but grow to values larger than unity outside
of the core regions \citep{tang13}.

This paper is organized as follows. Section \ref{section_obs} describes our ALMA observations. 
Maps of dust continuum, polarized emission and B-field morphologies are presented in Section \ref{result}.
We discuss the connection between polarization and B-field structures, together with a comparison to 
larger-scale data of the envelope of W51 e2/e8 and North from the SMA, in Section \ref{discussion}.
This section further debates B-field versus gravity with a new proposed measure.
Summary and conclusion are given in Section \ref{summary}.

\section{Observations}   \label{section_obs}

The project was carried out with the ALMA Band 6 receiver during the Early Science (Cycle 2, project "2013.1.00994.S").
Observations were done in three execution blocks (EBs) on July 18, 2015.
The three EBs were calibrated separately in flux, bandpass and gain.
Polarization calibration was performed after merging the three calibrated EBs
following the standard polarization calibration for ALMA\footnote{
CASA guide https:$//$casaguides.nrao.edu$/$index.php$/$3C286\_Polarization}.
A detailed analysis of the instrumental polarization in Band 6 is given in \citet{nagai16}
which conclude that linear polarization at a level of $<0.1$\% is detectable. 
The array included 38 antennas with (projected) baselines ranging from 13 m to 1492 m.
The four basebands were set in FDM mode (3840 channels for 1.875 GHz with a resolution of 488 kHz).
The calibration (bandpass, phase, amplitude, flux) was performed using CASA\footnote{http:$//$casa.nrao.edu$/$} v4.5.0. 
J1924-2914 and J1751+0939 were used for bandpass, and J1922+1530 for phase calibration. 
With a flux of 0.175 Jy at 232.9 GHz, J1922+1530 also provided the flux scale with the reference flux calibrator Titan. 
The polarization calibrator was J1924-2914,
which was measured to have a polarization fraction of 2.56\% and a polarization position angle of 45.6$\degr$,
in agreement with other ALMA measurements.
W51 was observed with three separate pointings, with phase centers on e2, e8 and North on 
$(\alpha, \delta)_{\rm{J}2000.0}$=(19:23:43.95, 14:30:34.00), (19:23:43.90, 14:30:27.00) and (19:23:39.95, 14:31:05.50), respectively. 
The images presented here are with natural weighting, which gives a synthesized beam resolution $\theta\sim0\farcs$28$\times$0$\farcs$26 with an orientation of 33$\degr$. 
The sensitivities (1$\sigma$) are 6 mJy/beam (e2, e8) and 1.4 mJy/beam (North) for Stokes $I$, 0.15 mJy/beam (e2, e8) and 
0.08 mJy/beam for Stokes $Q$, and 0.19 mJy/beam (e2, e8) and 0.10 mJy/beam for Stokes $U$
(Figure \ref{figure_composite_pol} and \ref{figure_composite_field}).
Since polarization measurements $I_p=\sqrt{Q^2+U^2}>0$ have a positive bias (while both $Q$ and $U$ can
be negative), we debias in the high signal-to-noise regime ($I_p\geq3\sigma_p$) with 
$I_p=\sqrt{Q^2+U^2-\sigma_{Q,U}^2}$, where $\sigma_Q\approx \sigma_U$ are the noise levels in $Q$
and $U$ \citep{leahy89, wardle74}.
For all results displayed in this paper we impose the two simultaneous conditions of having Stokes  $I\geq$3$\sigma$ 
and $I_p\geq$3$\sigma_p$.
Resulting median uncertainties and standard deviations of uncertainties for polarization percentages $p=I_p/I$ are
0.21\% and 0.15\% for W51 e2, 0.28\% and 0.16\% for W51 e8, and 0.16\% and 0.13\% for W51 North. 
Maximum and minimum uncertainties are 0.50\% and 0.02\%. 
Median uncertainties and standard deviations of uncertainties for the orientations of polarization position angles
are $2.5^{\circ}$ and $2.1^{\circ}$ for e2, $2.7^{\circ}$ and $2.1^{\circ}$ for e8, and $3.0^{\circ}$ and $2.4^{\circ}$ for North. Maximum uncertainties are around 9$^{\circ}$.

For comparison, we also present new larger-scale SMA maps combining data from the subcompact array 
configuration (described 
originally in \citet{tang13} for W51 North) and the compact array configuration (described originally in \citet{zhang14}
for W51 e2, e8, and W51 North). 
Additionally, unpublished data on W51 e2 and e8 from subcompact array observations are added.
The resulting images have $\theta\sim2\farcs13\times1\farcs88$ with an orientation of 28$\degr$, which 
captures the envelope scale and the 
previously unseen
connection between W51 e2 and e8 (panel (a) and (f) in Figure \ref{figure_composite_pol}
and \ref{figure_composite_field}).

\section{Results}        \label{result}

\subsection{230~GHz Dust Continuum}  \label{result_continuum}

The continuum emission is well detected and resolved at a 0$\farcs$26 resolution towards W51 e2, W51 e8 and W51 North (Figures \ref{figure_composite_pol} and \ref{figure_composite_field}).
The total detected continuum emission at 230 GHz in Stokes $I$ is 4.0, 3.7 and 6.8 Jy for W51 e2, e8 and North, respectively.

The W51 e2 core is resolved into 4 sub-cores (Table \ref{table_cores}). 
The flux densities of these dense cores are determined by two-dimensional Gaussian fits to be 
2.54, 1.04, 0.22 and 0.10 
Jy for W51 e2-E, e2-W, e2-NW and e2-N, respectively, and there is faint emission (0.2 Jy) in between these cores.
The W51 e2 core has also been revolved at 1.3 mm by \citet{shi10a} using the SMA with $\theta\sim1\farcs$1.
Their reported flux densities are 2.15$\pm$0.12, 0.62$\pm$0.12 and 0.73$\pm$0.08 Jy for W51 e2-E, e2-NW and w2-N, respectively.
Our newly reported flux densities are within the 3$\sigma$ uncertainty levels of the ones in \citet{shi10a}.
We note that this difference can be explained by the coarser angular resolution of the SMA observations with respect to the separations of these sub-cores. 
In addition, in our ALMA data, there is no detection at the location of W51 e2-N reported in \citet{shi10a}. Instead, there is a sub-core detected in the presented ALMA data 0$\farcs$5 west of W51 e2-N. We attribute this emission to W51 e2-N in \citet{shi10a}
and hence, the nomenclature is kept unchanged. 
This shift in position is a result of the interferometric filtering effect,  where the emission from relatively smooth and extended structures will be filtered out, and the structures revealed by ALMA have fewer artefacts due to a more complete uv-coverage.

The W51 e8 region is resolved into two cores, e8-N and e8-S, with some additional faint emission in the south of 
e8-S. The flux densities are 2.65 Jy and 0.34 Jy for the e8-N and e8-S core, respectively.

The continuum emission towards the W51 North region shows several cores aligned in an east-west direction. 
These cores have been resolved and reported in \citet{tang13}.
The SMA2 core is now further resolved into a new core to its southwest (N2) and a likely additional emerging peak to 
its east (SMA2-E).
Hereafter, W51 N1 refers to SMA1, W51 N2 to the newly resolved peak southwest of SMA2, W51 N3 to SMA3 and W51 N4 to SMA4.
The flux densities are 2.88, 1.49, 0.96, 0.89, and 0.59 Jy for N1, SMA2-E, N2, N3, and N4, respectively
(Table \ref{table_cores}).

\subsection{Polarization}  \label{result_polarization}    

\begin{figure}

\includegraphics[width=14.0cm]{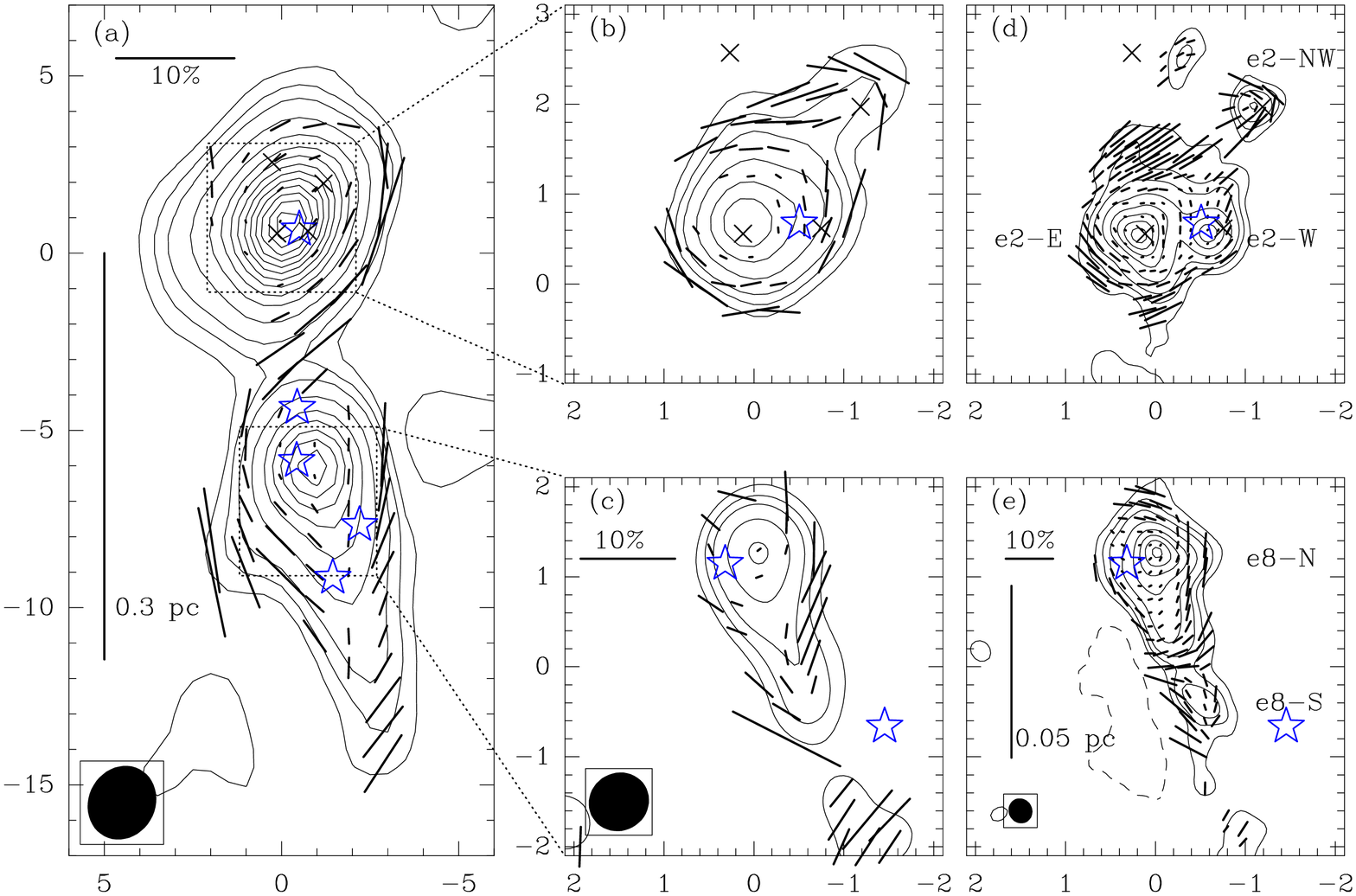}
\includegraphics[width=14.0cm]{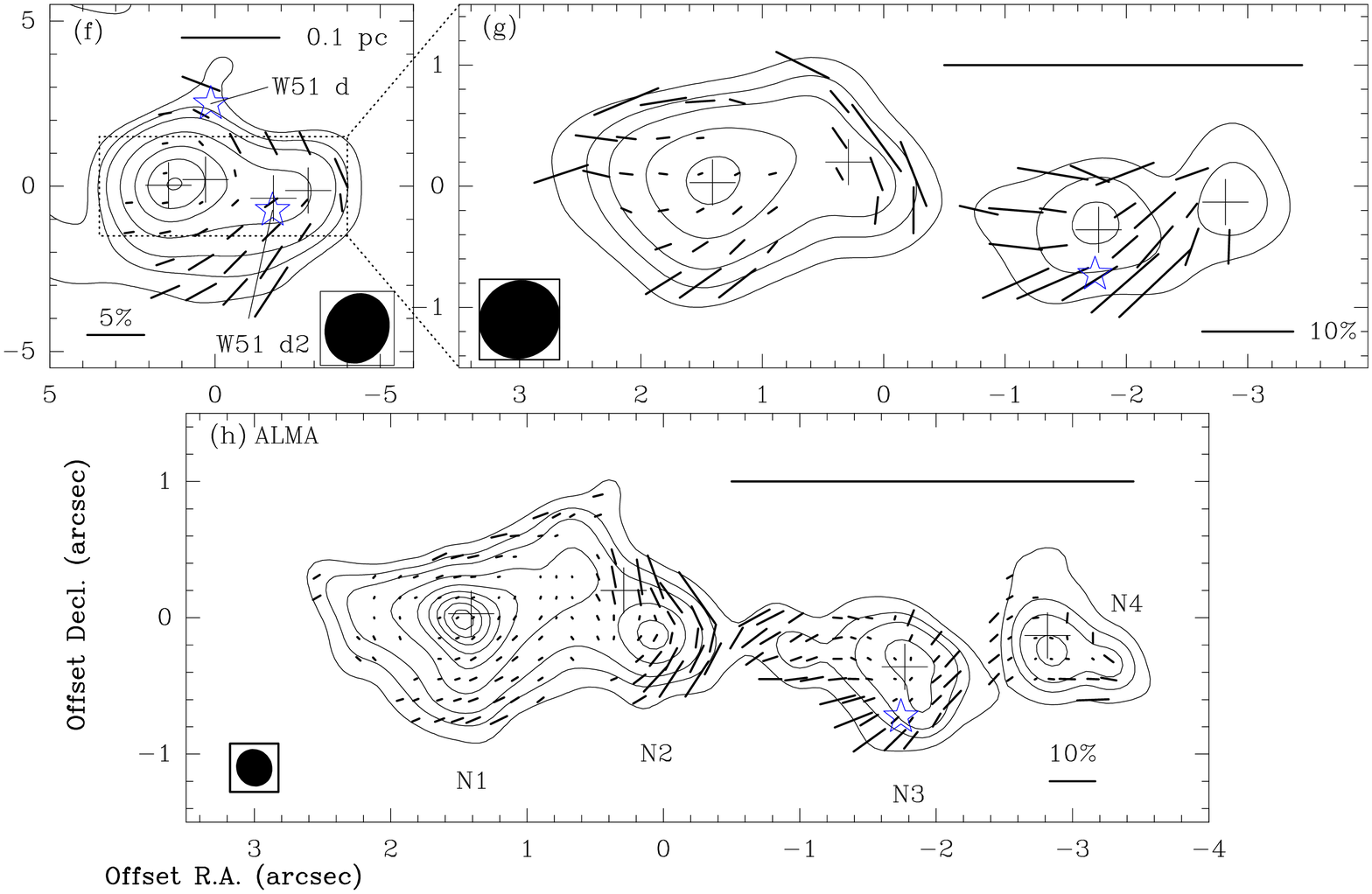}
\caption{\tiny 
Polarization maps of W51 e2, e8, and W51 North with various angular resolutions from ALMA and the SMA. 
Contours are Stokes $I$ dust continuum intensity.
Polarization orientations are displayed with segments with their lengths scaled with polarization percentage $p=I_p/I$.
Contours are 3, 6, 10, 20, 35, 50, 65, 80, 95...$\times\sigma$, where $1\sigma$ is 75 mJy/beam in panel (a), 60 mJy/beam in panel (b) and (c), 
6 mJy/beam in panel (d) and (e), 140 mJy/beam in panel (f), and 90 mJy/beam in panel (g).
Contours in panel (h) are identical to panel (d) and (e) but $1\sigma$ is 1.4 mJy/beam.
Panel (a) and (f): SMA observations with $\theta\sim 2\arcsec$ at 345~GHz probing larger envelope scales, revealing the connection between W51 e2 and e8 in panel (a) and the W51 North region in panel (f).  
Panel (b), (c), and (g): SMA observations with $\theta\sim0\farcs7$ at 345 GHz towards W51 e2 in panel (b), 
W51 e8 in panel (c) (images adopted from \citet{tang09b}), and W51 North in panel (g) (image adopted from \citet{tang13}).  
Panel (d), (e), and (h): ALMA observations in Band 6 at 230~GHz with $\theta\sim 0\farcs26$ towards W51 e2 in panel (d),
 W51 e8 in panel (e), and W51 North in panel (h).   
Crosses in the panels (a), (b) and (d) mark the known sub-mm sources W51 e2-E, e2-W, e2-NW and e2-N, counter-clockwise around the continuum peak.
Pluses in the panels (f), (g), and (h) mark the known sub-mm sources SMA1, SMA2, SMA3 and SMA4 from east to west.
N1 to N4 label the clearly resolved peaks in the ALMA observations.
Blue stars indicate UCHII regions. Synthesized beams for each observation are shown with black filled ellipses. Polarization segments are gridded to and displayed at half of the synthesized beam resolution. 
}
\label{figure_composite_pol} 
\end{figure}

\begin{figure}
\includegraphics[width=9.5cm]{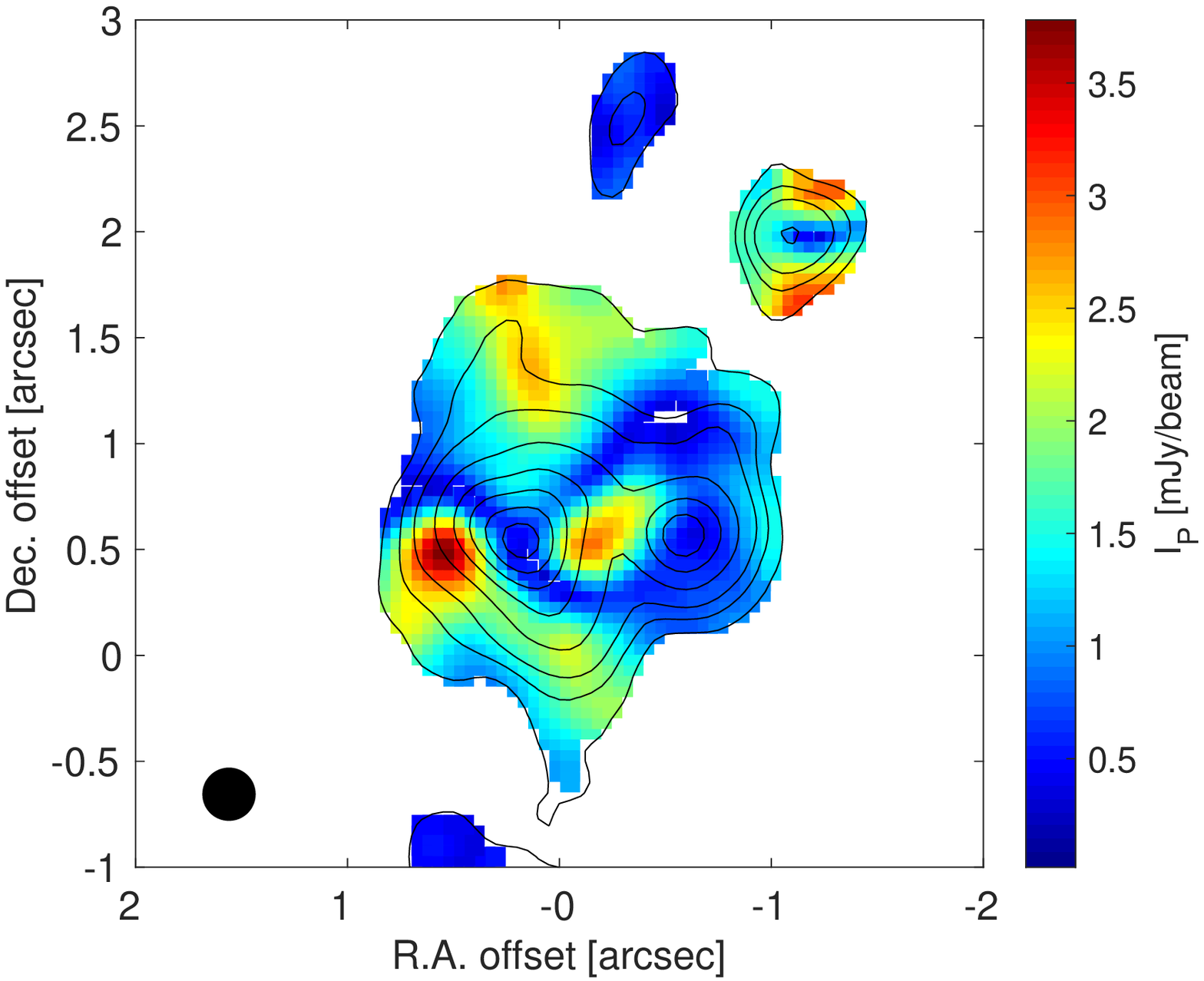}
\includegraphics[width=9.5cm]{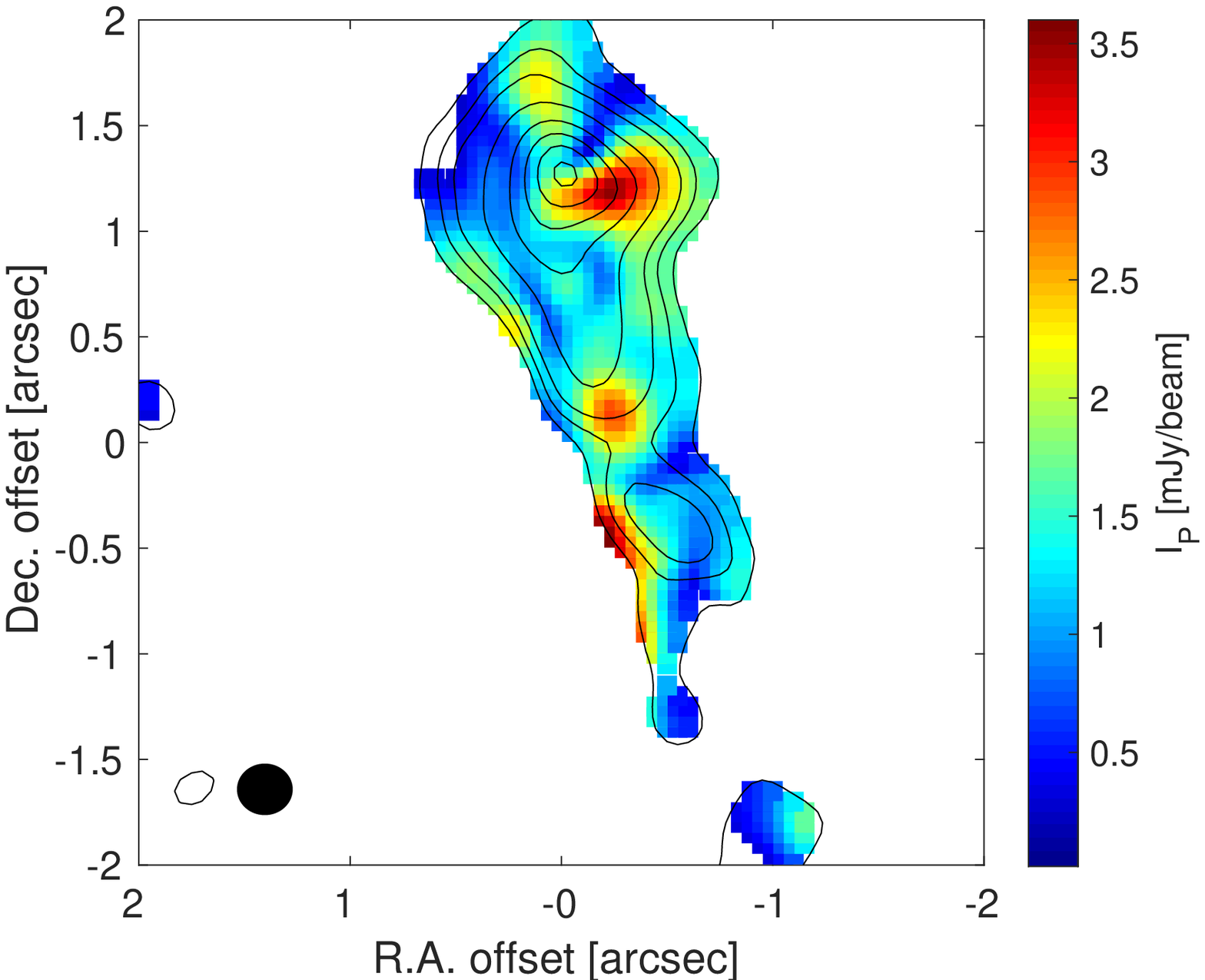}
\includegraphics[width=16cm]{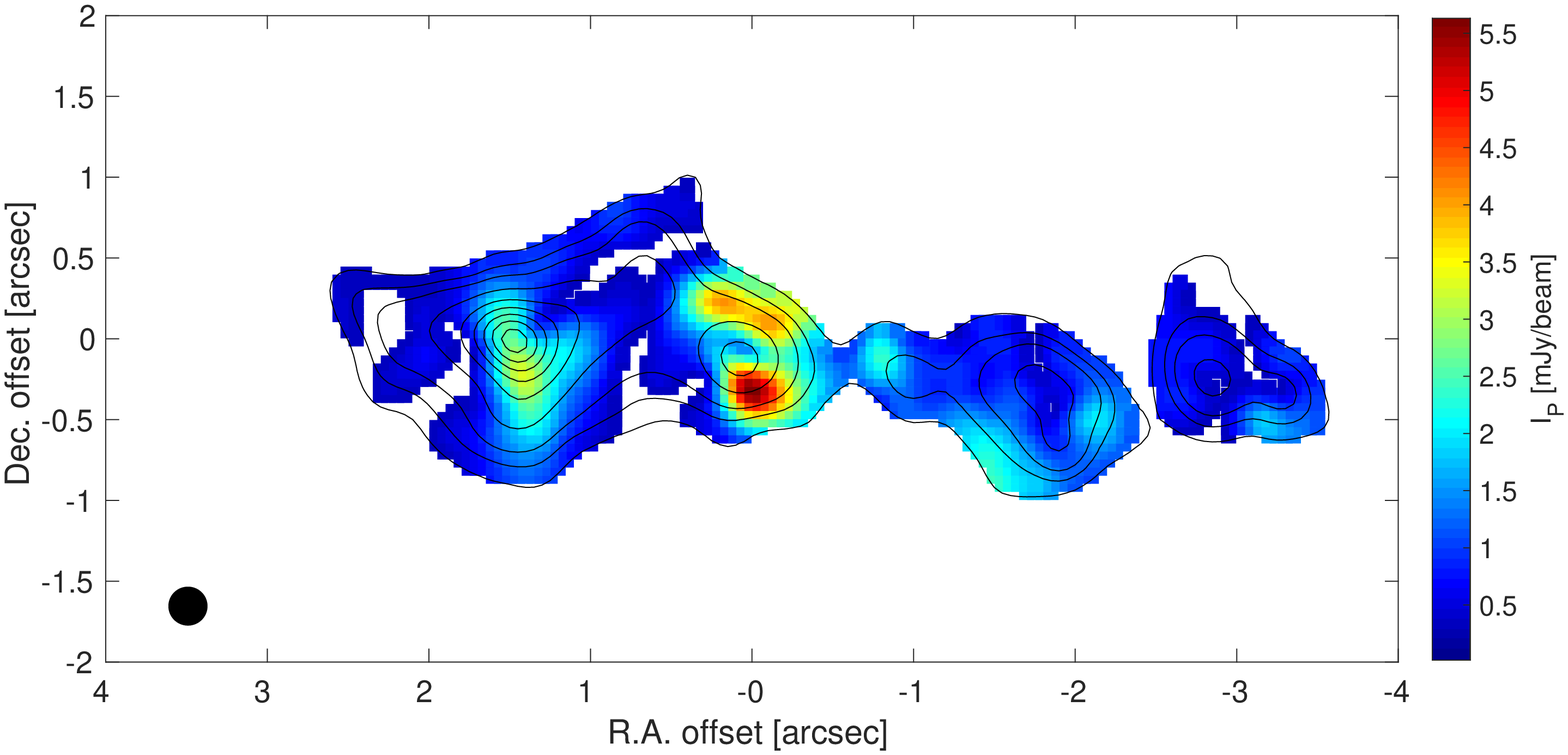}
\caption{Polarized emission $I_p$ for e2 (top left), e8 (top right) and North (bottom).
Contours are dust continuum 
with levels as described in Figure \ref{figure_composite_pol},
color scale is in units of mJy/beam for $I_p$.
Note: $I_p$ is reproduced from the ALMA maps in Figure \ref{figure_composite_pol} and for a better visual impression and display of features, the data are additionally overgridded and shown at five times the 
synthesized beam resolution. 
Synthesized beams are shown with black filled ellipses.}
\label{figure_I_pol} 
\end{figure}

Polarized emission above 3$\sigma$ is seen and resolved with ALMA
at 230~GHz
 in W51 e2, e8, and W51 North 
(panel (d), (e), and (h) in Figure \ref{figure_composite_pol}).
Polarization holes or de-polarization zones in the earlier observations with the SMA
at 345~GHz
($\theta\sim 0\farcs7$, here reproduced from \citet{tang09b} in panel (b), (c), and (g) in 
Figure \ref{figure_composite_pol}) are now resolved with ALMA.
It is obvious from Figure \ref{figure_I_pol} that the {\it absolute polarized emission} $I_p$ does not simply scale with
Stokes $I$. W51 e2, e8, and W51 North show high- and low-emission zones and spots in $I_p$ that appear to have no counterparts in $I$.  This is clearly seen in the plots $I_p$ versus $I$ 
(Figure \ref{figure_all_correlations_polarizations}, Appendix) that show
no correlation but a broad scatter between these two observables. 
We note that this joint interpretation of $I$ with $I_p$ assumes that both the total intensity and 
the polarized signal result from the same structure along the line of sight within a synthesized beam 
resolution. This means that no significant contamination from background, foreground, or any 
intervening structures should be present. 

{\it W51 e2:} 
Zones of decreasing and minimum polarized emission are centered on e2-W,  across e2-E along a 
northeast-southwest direction, and away from the e2-E emission peak along a narrow straight line 
towards northwest (Figure \ref{figure_I_pol}).  A peak is detected east of e2-E and in 
between the east and west core. 
The e2-NW satellite core reveals itself with a stripe of minimum polarization along an east-west
direction, a growing signal towards the east and two maxima in the north and south, 
displaying an almost perfect north-south symmetry. 
This symmetry is also reflected in the magnetic field morphology (Section \ref{result_magnetic_field}, 
Figure \ref{figure_composite_field}).

{\it W51 e8:}
Absolute polarized emission peaks are seen west of the e8-N peak, in between e8-N and e8-S, and east of e8-S at 
the lowest Stokes $I$ emission contour.  Zones of minimum polarization appear east of the e8-N peak 
and to its north-west. Both the northern and southern end of e8-S show low-level absolute polarization.

{\it W51 North:}
Peaks in absolute polarized emission are detected north and south of the core N2 around R.A. offset 0. 
To a smaller extent, two additional 
local maxima are seen north and south of the core N1 around R.A. offset $1.5$.  
The remaining cores and connections in between them are 
mostly weakly polarized around 1 mJy/beam or less, with a few patches that are slightly more polarized up to about 
2 mJy/beam. 

\begin{figure}
\includegraphics[width=9.5cm]{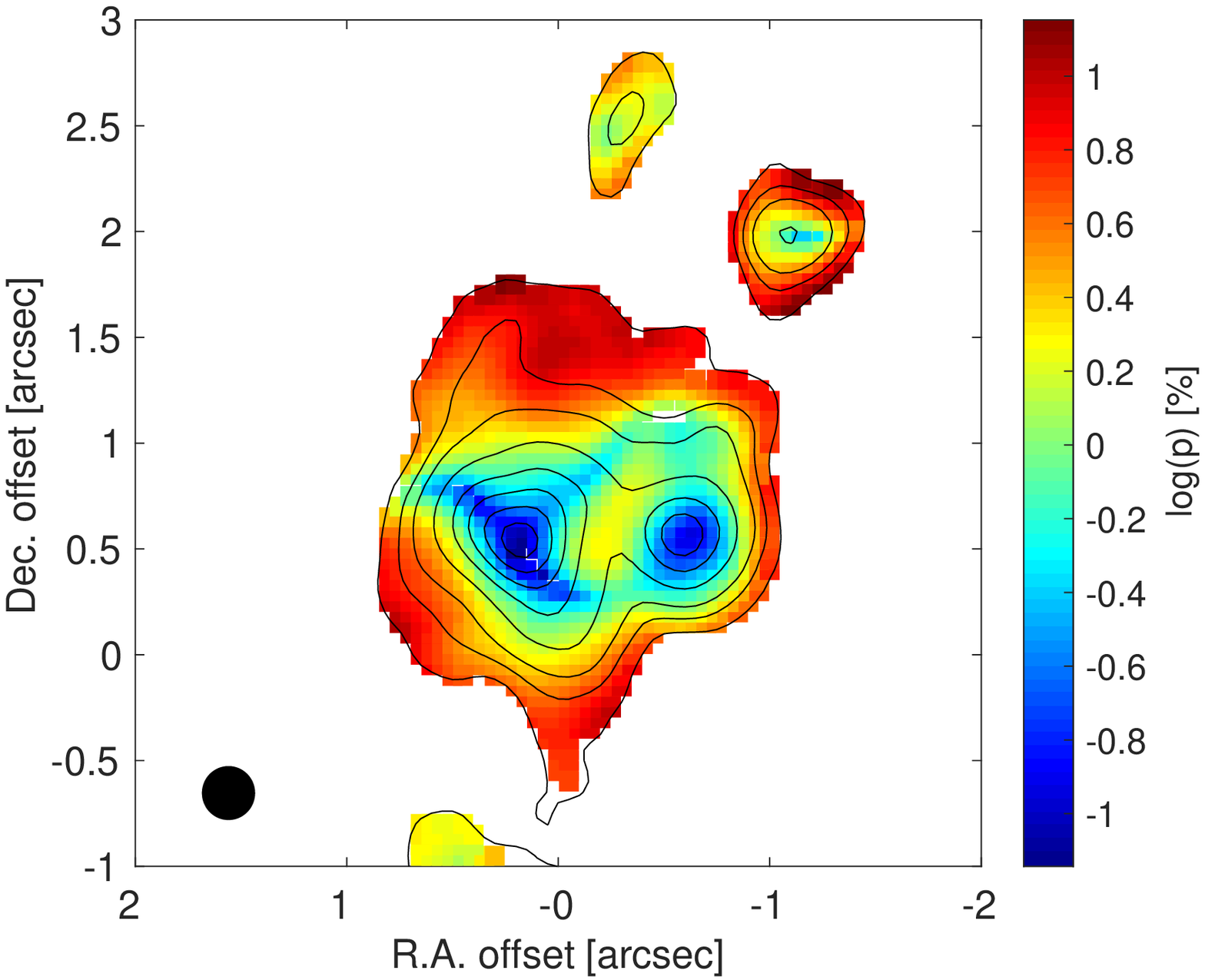}
\includegraphics[width=9.5cm]{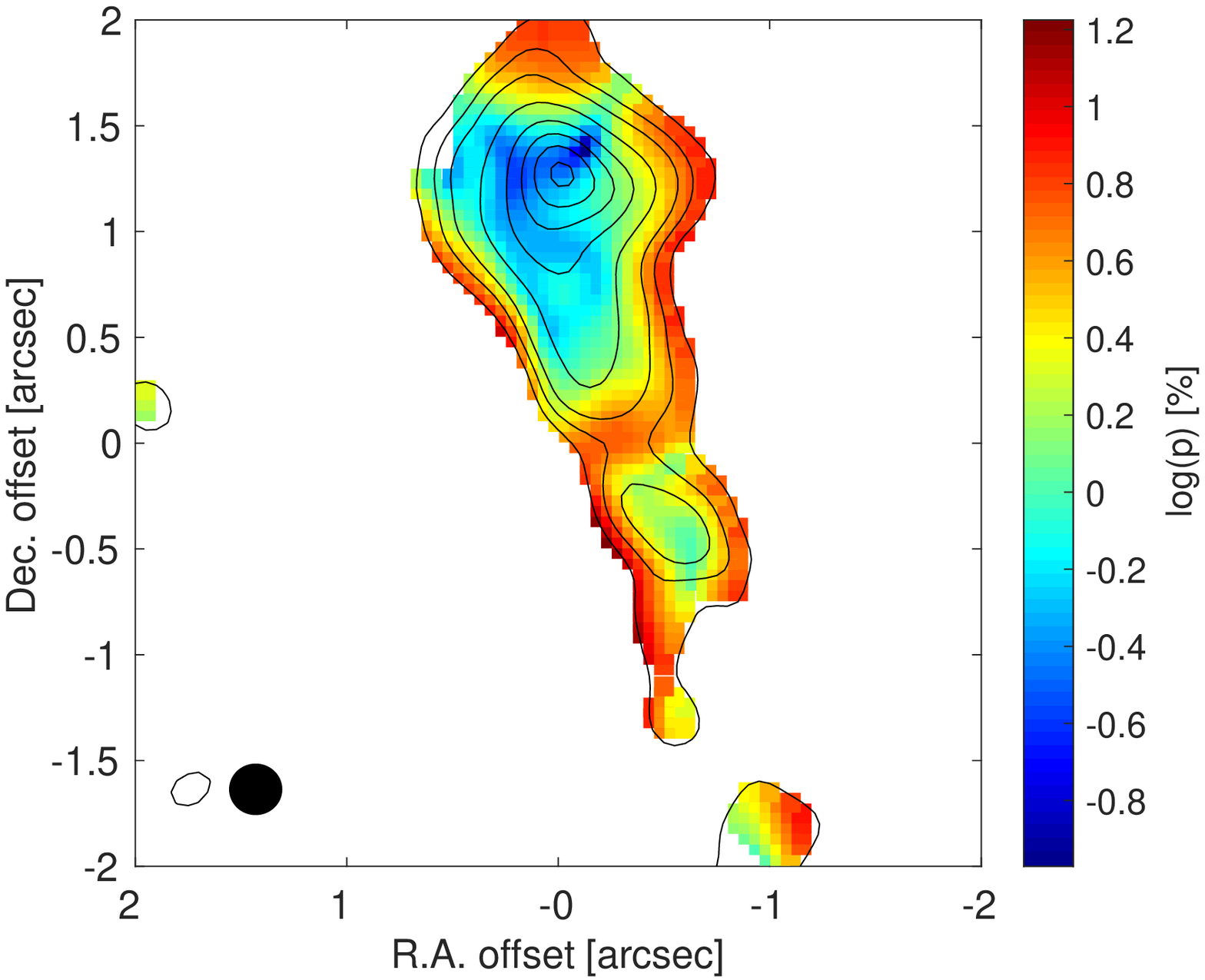}
\includegraphics[width=16cm]{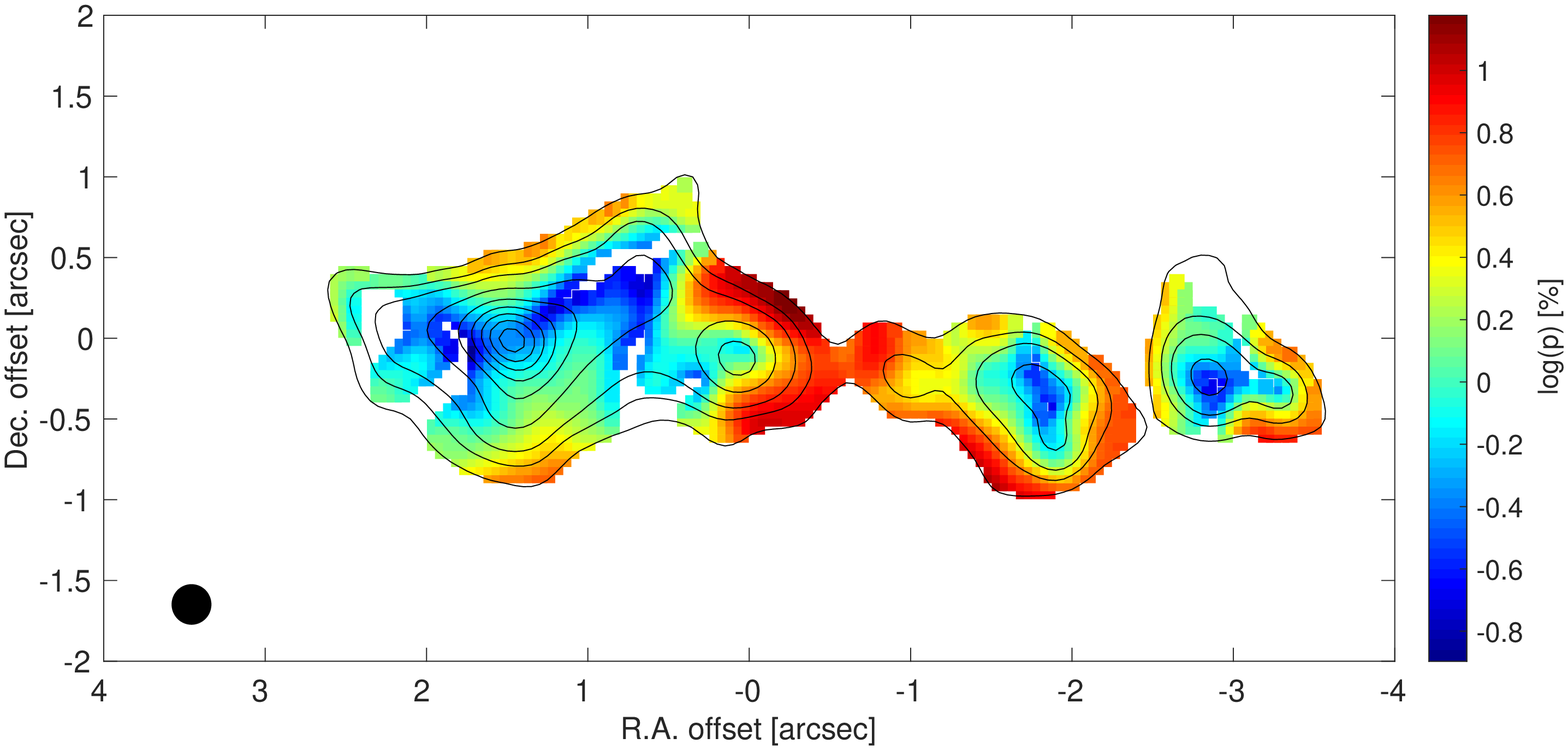}
\caption{Polarization percentages $p=I_p/I$ for e2 (top left), e8 (top right) and North (bottom).
Color scale is $\log(p)$.
Note: Identical to Figure \ref{figure_I_pol}, the data are overgridded and shown at five times the 
synthesized beam resolution.
$p$ is extracted from the ALMA maps in Figure \ref{figure_composite_pol}
where it is encoded in the lengths of the polarization segments.
Synthesized beams are shown with black filled ellipses.}
\label{figure_pol_perc} 
\end{figure}

\begin{figure}
\includegraphics[width=9cm]{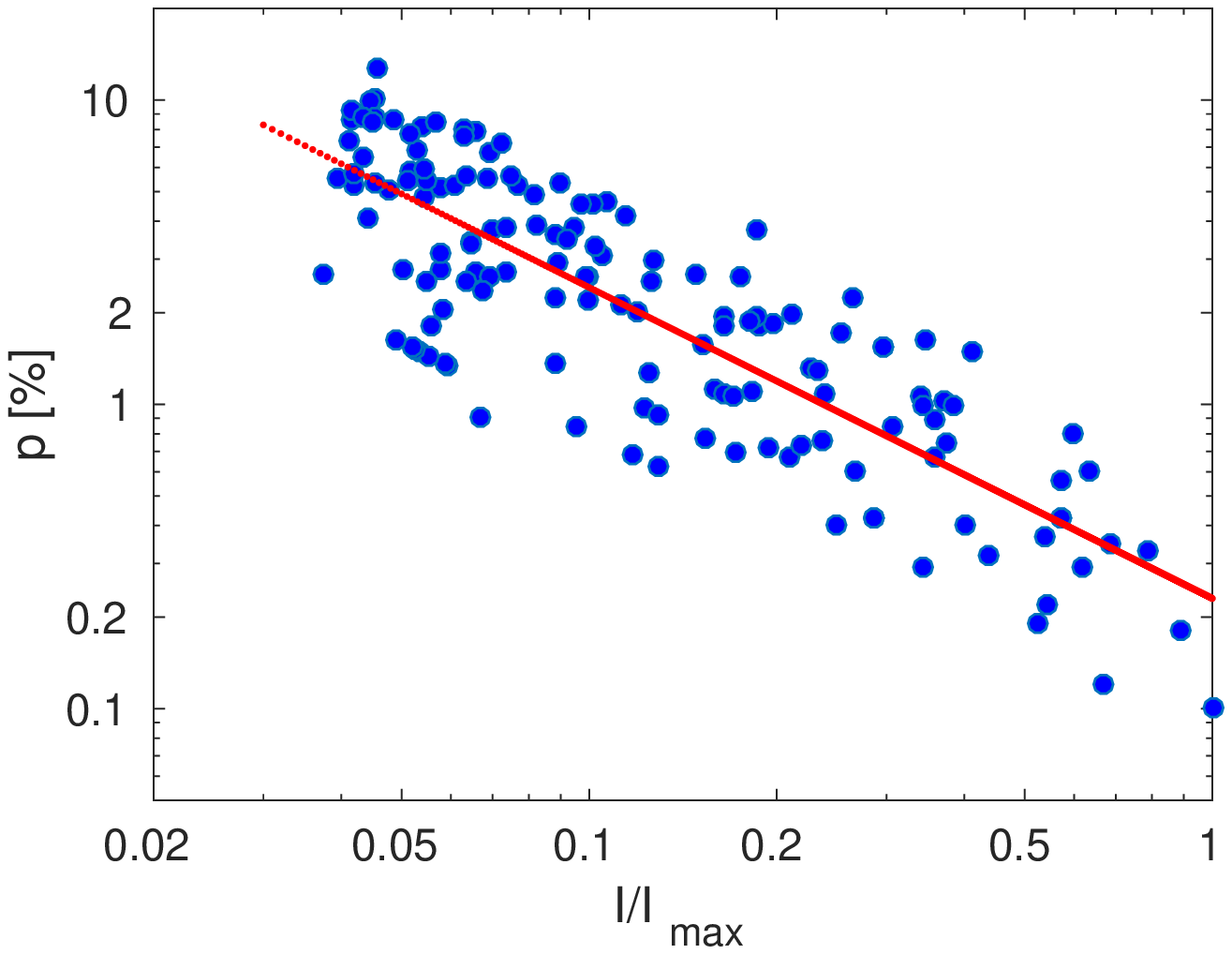}
\includegraphics[width=9cm]{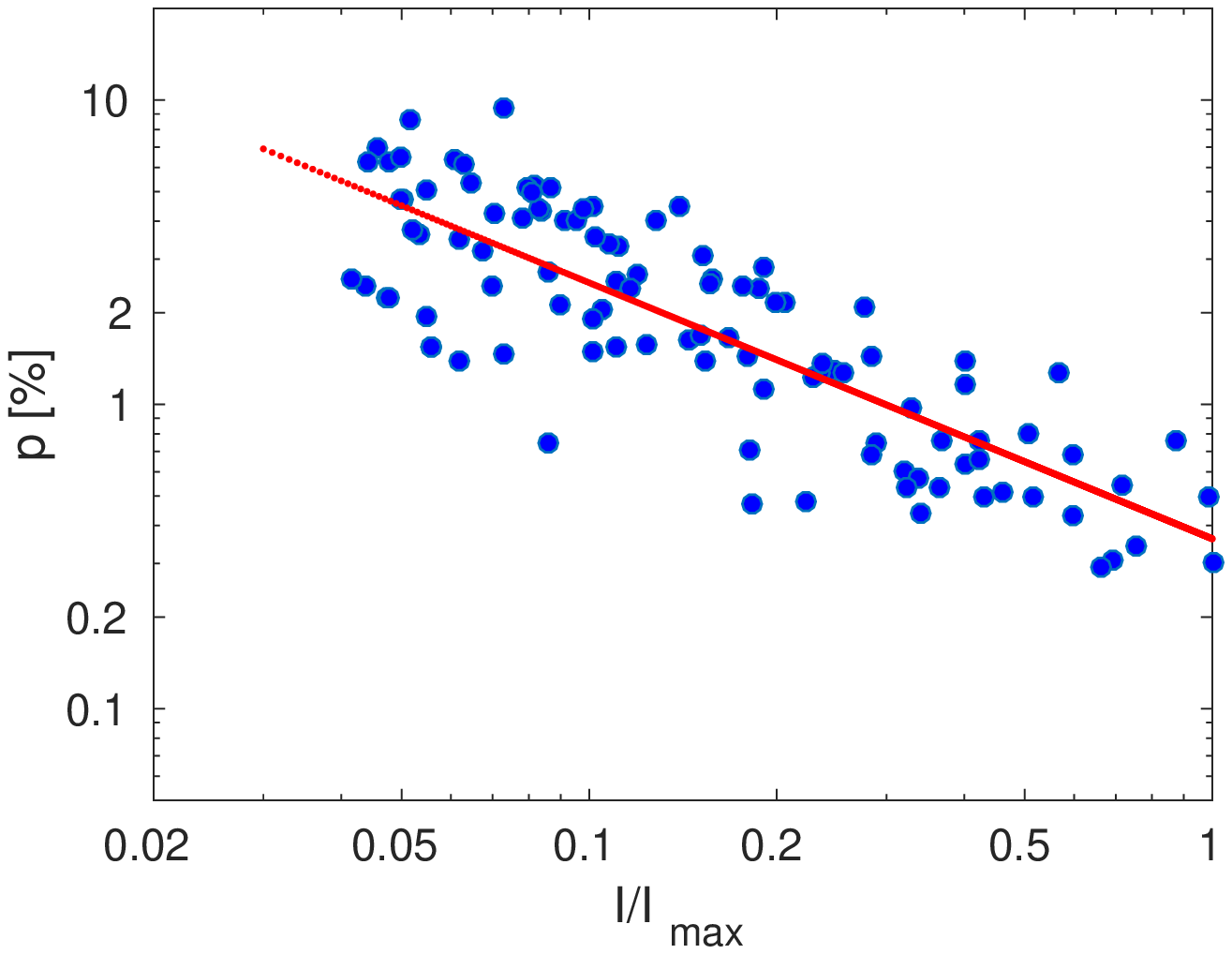}
\includegraphics[width=9cm]{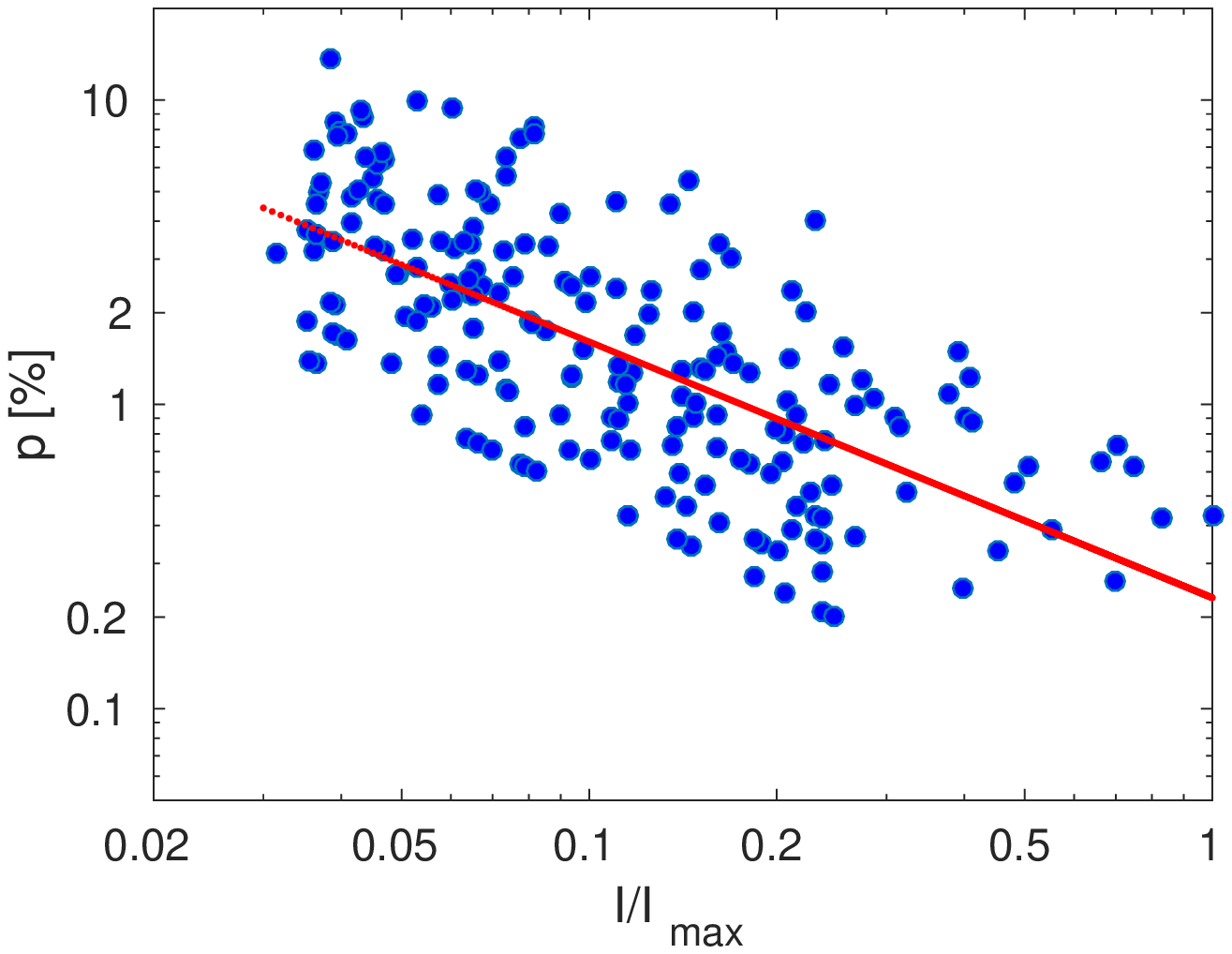}
\includegraphics[width=9cm]{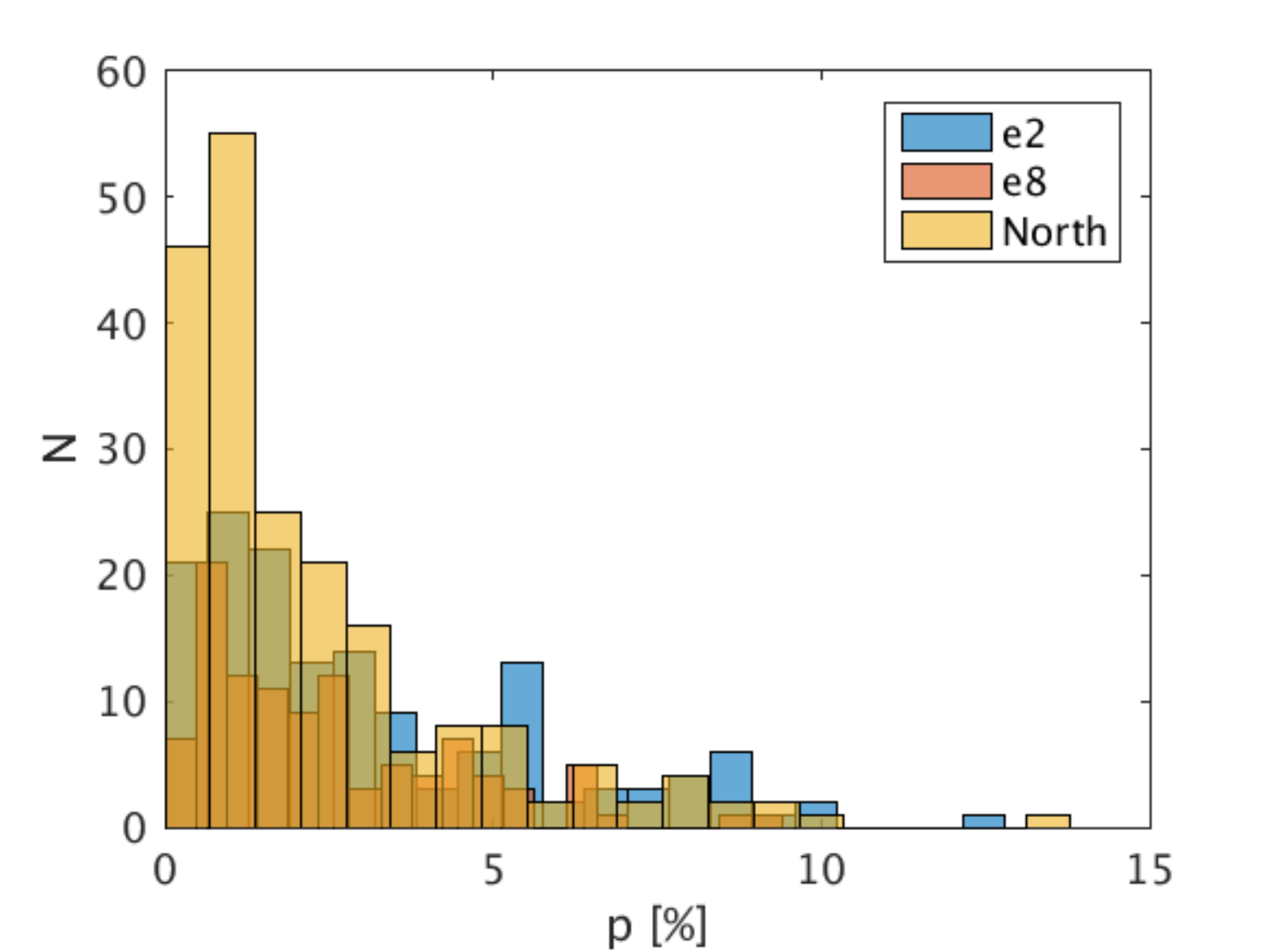}
\caption{Polarization percentage $p=I_p/I$ versus Stokes $I$, normalized to $I_{max},$ for e2 (top left), e8 (top right), and North (bottom left).
Unlike the overgridded data displayed in Figure \ref{figure_pol_perc}, the data here are extracted from maps 
gridded to only half of the 
synthesized beam resolution (panels (d), (e), and (h) in Figure \ref{figure_composite_pol}).
The red solid lines are the best-fit power laws with indices -1.02 (e2), -0.84 (e8), and -0.84 (North).
Bottom right panel: histograms of polarization percentages $p$.
Averages and standard deviations are 3.1\% and 2.2\%, 2.5\% and 2.0\%, and 2.3\% and 2.3\% for e2, e8, and North.
Maximum and minimum polarizations are 13\% and 0.1\% (e2), 9\% and 0.3\% (e8), and 14\% and 0.2\% (North). }
\label{figure_I_perc} 
\end{figure}

Unlike the above {\it absolute polarization}, the {\it relative polarization} $p=I_p/I$ shows systematic trends where
$p$ grows with decreasing Stokes $I$ (Figures \ref{figure_composite_pol}, \ref{figure_pol_perc}). This is the typically observed anti-correlation of {\it relative polarization} versus Stokes $I$, i.e., $p$ versus $I/I_{max}$ if normalized by the maximum Stokes $I$ value,  as e.g.,  in \citet{tang09b}.
Without any exception, all cores in W51 e2, e8 and North show local minima in $p$ at their Stokes $I$ emission peaks.
Similarly, maximal polarization percentages are always associated with the lowest contours in $I$. Nevertheless,  
a constant $I$ contour can show a significant variation in relative polarization (Figure \ref{figure_pol_perc}).  
This means that symmetries in $I$
are not necessarily preserved in $p$, as e.g., evident in e2-E and N2.
This naturally leads to a scatter in the $p$ versus $I$ correlation.  This scatter is relatively broad over
almost one order of magnitude (Figure \ref{figure_I_perc}). It, thus, likely hints a dependence on additional 
physical parameters that are not captured in this simple anti-correlation. 
Polarization percentages go over two orders of magnitude, ranging from around 0.1\% to 10\% in all 
sources (Figure \ref{figure_I_perc}). 
The anti-correlation can be fit with power-laws with indices -1.02 (e2), -0.84 (e8), and -0.84 (North). 
Similar slopes are seen in the larger-scale SMA 
345~GHz (850~$\mu$m)
data (Section \ref{discsussion_larger_scale}) and in an 
SMA-BIMA comparison for e2/e8 \citep{tang09b} 
and in a comparison with CSO 
(at 350~$\mu$m)
and JCMT 
(at 850~$\mu$m)
observations for North \citep{tang13}.
A possible connection between the spatially varying polarization percentage $p$ and the observed
B-field morphologies is discussed in Section \ref{discussion_polarization}.

\subsection{Magnetic Field Morphologies}  \label{result_magnetic_field}

\begin{figure}
\includegraphics[width=14cm]{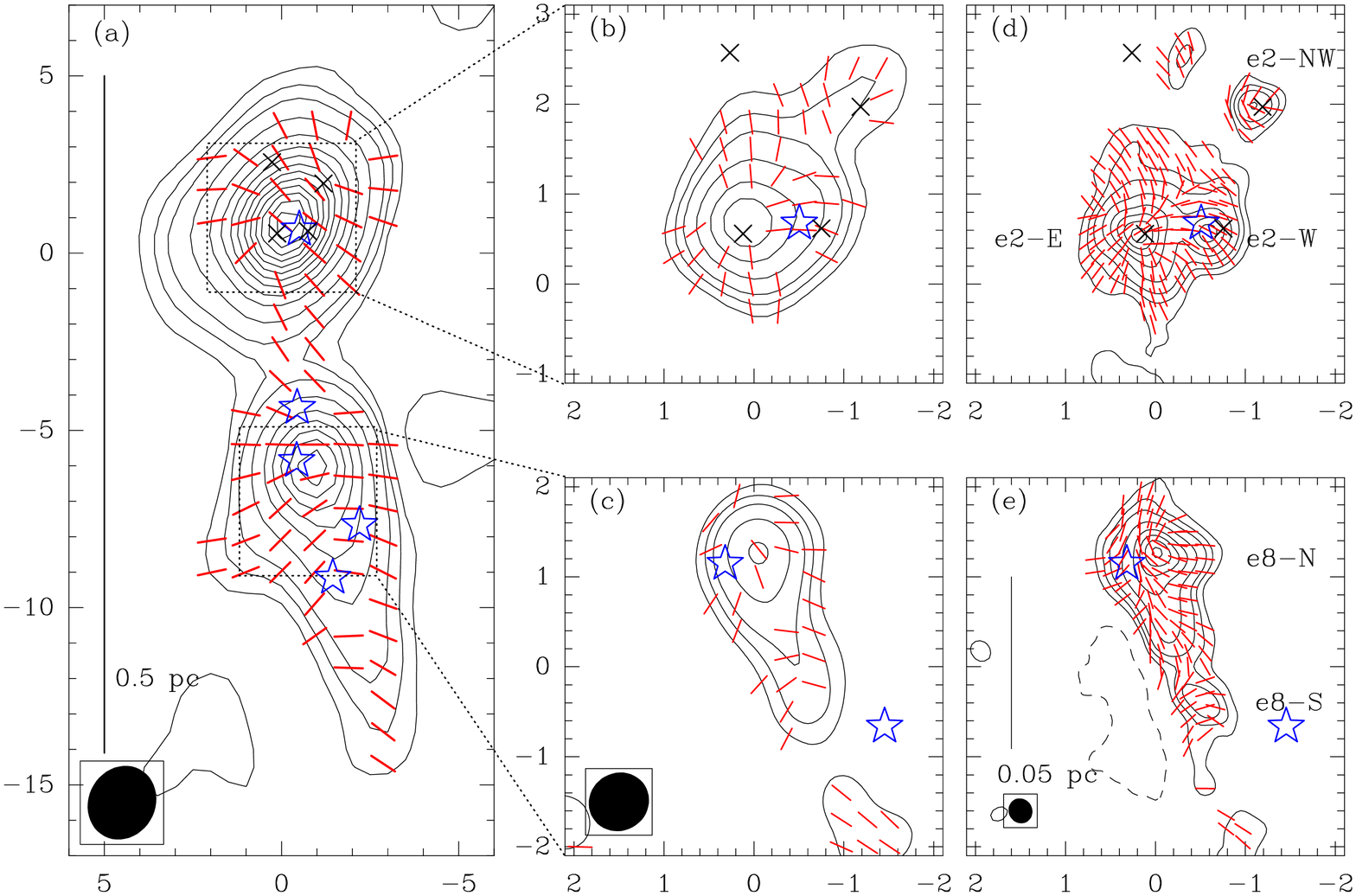}
\includegraphics[width=14cm]{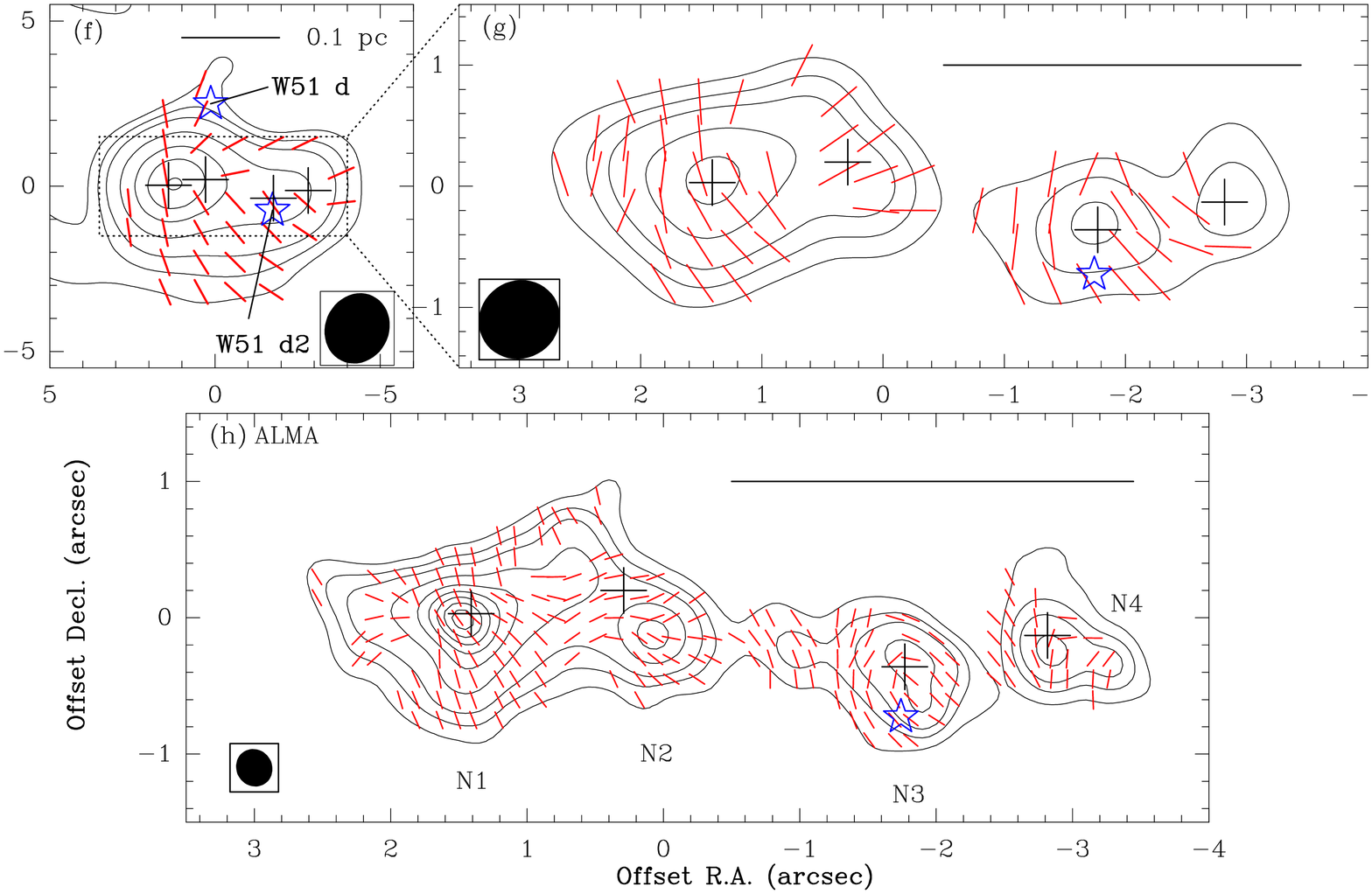}
\caption{\scriptsize 
Identical to Figure \ref{figure_composite_pol} but with magnetic field orientations shown with red segments. 
B-field orientations are rotated by 90$^{\circ}$ with respect to the detected polarization orientations in 
Figure  \ref{figure_composite_pol}.}
\label{figure_composite_field} 
\end{figure}

Magnetic (B) field morphologies are clearly detected, revealing organized, coherent and connected
structures. Furthermore, substructures in and in between individual cores, and shaped B-field morphologies 
in satellite cores can now be identified (Figure \ref{figure_composite_field}).
In this section, B-field orientations are generated by rotating the originally detected polarization orientations in 
Figure \ref{figure_composite_pol} by 
$90^{\circ}$. B-field segments are all displayed with equal lengths, neglecting information
about relative polarization (Section \ref{result_polarization}).

{\it W51 e2:}
Overall, most of the B-field segments are pointing towards the main emission peak e2-E 
(panel (d) in Figure \ref{figure_composite_field}).  
In the closer vicinity of the e2-E peak, the segments are becoming almost radial-like. 
Overall, the field structure around e2-E resembles a dragged-in morphology.
Around e2-W, the field segments in the south-western peripheral area are bent towards the e2-W
emission peak, while the remaining surrounding segments still point towards the main peak e2-E, 
leaving straight segments along an east-west direction between e2-E and e2-W.  This possibly indicates
that e2-W is pulled towards the more massive e2-E core. 
The main core e2-E displays two additional features. Firstly, the previous depolarization stripe along the 
northeast-southwest direction in the SMA observation (panel (b) in Figure \ref{figure_composite_field})
is resolved, clearly showing a continuation of B-field segments that now appear to converge from above
and below towards a mid-plane along this stripe. Secondly, perpendicular to this, B-field segments align
along a straight northwest-southeast axis. This is particularly obvious in the northern upper plane with 
field segments converging symmetrically from both east and west towards this central straight axis. 
Finally, the satellite core e2-NW appears with a cometary- or bow-shock-shaped B-field morphology.
This core hints a pinched field structure in the west and a curved bow-shock structure in the east
with a north-south symmetry. These features might suggest that e2-NW is passing through the 
ambient (lower-density) medium from west to east.

{\it W51 e8:} 
The more elongated e8 structure is clearly detected and further resolved into the main 
core e8-N and e8-S (panel (e) in Figure \ref{figure_composite_field}).  
The polarization coverage is significantly improved as compared to the 
earlier SMA map (panel (c) in Figure \ref{figure_composite_field}). While the western side 
of e8-N displays B-field segments that appear oriented towards the emission peak, the eastern
and particularly the northeastern side indicate field lines that are bending away, more closely aligning
with a north-south axis. 
The smaller core e8-S hints a cometary field morphology. Although not as obvious
as e2-NW, it hints identical features with possibly pinched field lines in its southern end and more 
curved cometary-shaped segments in the northern part. This is suggestive of e8-S being pulled
north towards the more massive e8-N.  This impression is further supported by the possibly streamlined
B-field segments in the lower density bridge between the northern tip of e8-S and the southern tail of e8-N.
This morphology -- likely shaped by a streaming motion -- might also be present at the northern and northeastern
zones of e8-N, possibly indicating that the entire e8 is pulled north towards the more massive e2.

{\it W51 North:} 
This region harbours at least six cores, aligned along an east-west axis and almost uniformly spaced.  
Each core displays an organized magnetic field structure (panel (h) in Figure \ref{figure_composite_field}).
The two most massive cores -- N1 and N2 from east
to west around R.A. offsets 1\farcs5 and 0\arcsec -- both exhibit B-field segments oriented towards their emission peaks. 
While N1 appears with a clearly pinched and complete hourglass B-field morphology, symmetric around a 
northwest-southeast axis,  N2 only presents likely dragged-in field lines at its western end.  In the east, 
the field segments appear to open up again, be more straight and possibly oriented towards the larger and 
more massive N1. 
Similar to e2-W 
and e8-S, this characteristic field morphology might be symptomatic
for the less massive core (here, N2) being pulled towards the next more massive gravitational center (N1).
In contrast to N1 and N2, N3 around R.A. offset $-2\arcsec$ 
clearly reveals field orientations that are dominantly not oriented towards its
emission peak, except in the southwestern zone. Many of the B-field segments are largely north-south oriented  
but with a twist towards the east.  This is particularly noticeable in the eastern extension of N3 that forms a 
bridge (around R.A. offset $-1\arcsec$) to N2 and where probably another core is embedded.  
This overall bending of the entire group of 
field segments could again indicate that N3, as a whole, is being dragged to the more massive N2.
Finally, the smallest and least massive core N4 at the western end of W51 North (around R.A. offset $-3\arcsec$) 
possibly also shows  
a glimpse of field segments being oriented towards the next more massive N3 to the east, although 
some segments in the west also show a north-south alignment and some tendency towards the emission 
peak.

\section{Discussion}    \label{discussion}

\subsection{Polarization Structures and Magnetic Field Morphologies}  \label{discussion_polarization}

\begin{figure}
\includegraphics[width=9.5cm]{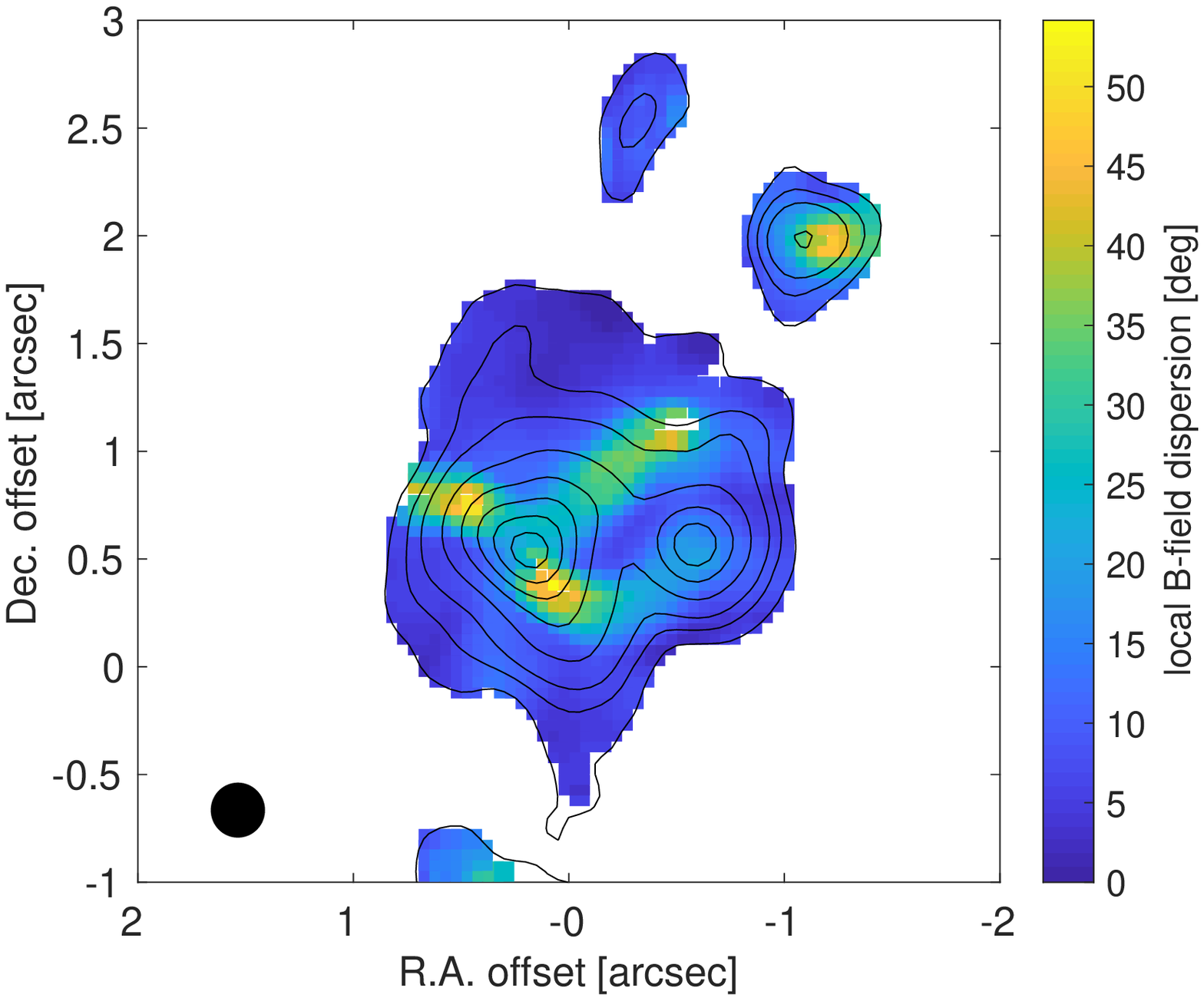}
\includegraphics[width=9.5cm]{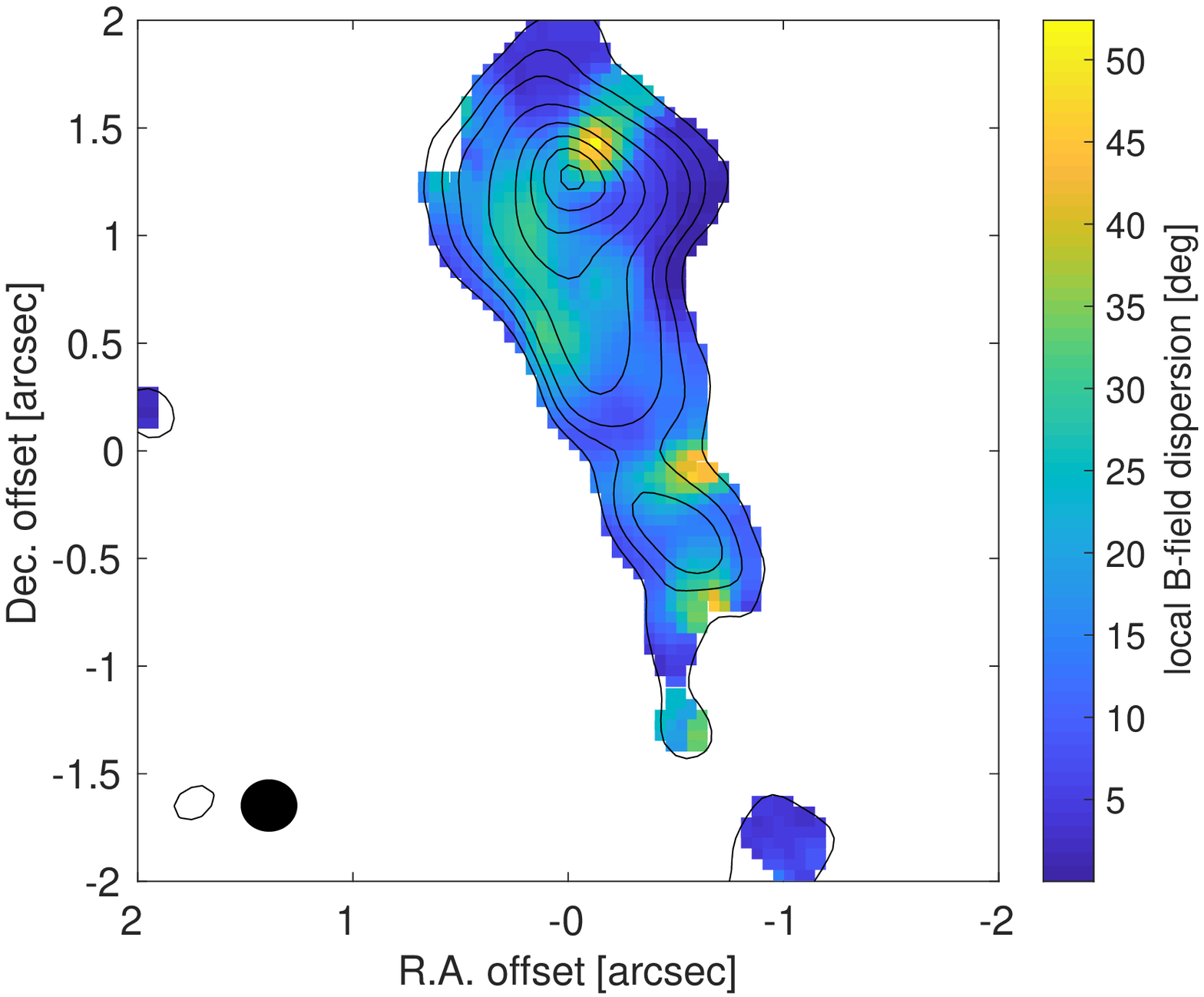}
\includegraphics[width=16cm]{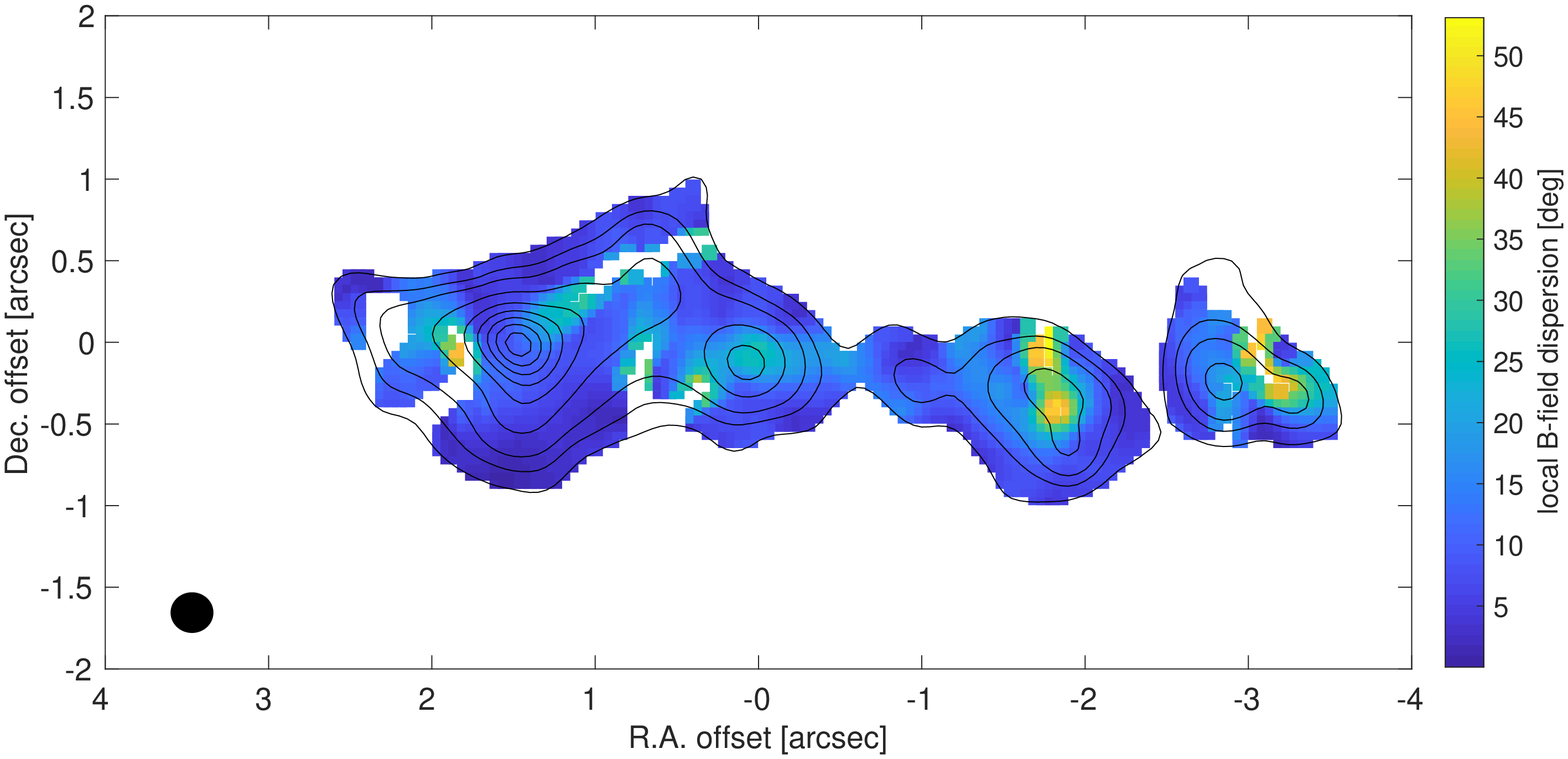}
\caption{Local magnetic field dispersion $\mathcal{S}$ (color scale in units of degrees, for radius $r_{disp}=0\farcs2$) overlaid on dust Stokes I contours for e2 (top left), e8 (top right), and North (bottom).
Note: Identical to the Figures \ref{figure_I_pol} and \ref{figure_pol_perc}, for a better visual impression, the data are overgridded and shown at five times the 
synthesized beam resolution. 
Synthesized beams are shown with black filled ellipses.}
\label{figure_field_dispersion} 
\end{figure}

The polarized emission -- both absolute and relative to Stokes $I$ (Figures \ref{figure_I_pol} and \ref{figure_pol_perc}) --
is clearly not random but appears organized, though in a non-trivial way.
Can this emission be understood together with the plane-of-sky projected B-field morphologies? 
Here, we explore correlations among the observables Stokes $I$, polarized emission $I_p$, polarization 
percentage $p=I_p/I$, and the local B-field dispersion $\mathcal{S}$. 
The latter one 
was recently probed on large data sets by {\it Planck}
\citep{planckXIX, planckXX} and BLASTPol \citep{fissel16}. 
The local B-field dispersion is defined as
\begin{equation}
\mathcal{S}(\mathbf{r},r_{disp}) = \sqrt{\frac{1}{N}\sum_{i=1}^N \left[PA(\mathbf{r})-PA(\mathbf{r+r_{disp,i}})\right]^2},
\end{equation}
where $PA$ is the position angle of an observed B-field segment at a location $\mathbf{r}$ and $i$ is 
counting neighboring B-field segments within an annulus $r_{disp}\ge|\mathbf{r_{disp,i}}|$ centered on $\mathbf{r}$.
Figure \ref{figure_field_dispersion} shows $\mathcal{S}$-maps, evaluated for 
$r_{disp}=0\farcs2$, which measures the field dispersion in an area slightly larger than the synthesized beam
$\theta\sim0\farcs$26. This means that $\mathcal{S}$ is capturing by how much a local field orientation
changes with respect to its nearest four and next-nearest four neighbors. $\mathcal{S}$ will select zones and display larger
values where the B-field bends more rapidly or changes orientation abruptly.  
In W51 e2-E (Figure \ref{figure_field_dispersion}, upper left panel) 
the northeast-southwest mid-plane, towards which the field segments seem to converge from north and south, 
is clearly identified with larger dispersion values $\mathcal{S}$.  Except this stripe, e2-E shows mostly small
values, which is a consequence of its radial-like field morphology.  
An additional zone with clearly enhanced $\mathcal{S}$-values
is in the northern low-emission region between e2-E and e2-W. This is again a zone where field segments
converge symmetrically from east and west
towards a central straight axis (Section \ref{result_magnetic_field}).
The $\mathcal{S}$-parameter also identifies 
the western side of the mid-plane in the satellite core e2-NW as a large-dispersion area. 
Similarly to e2-E, this reflects the mirror-symmetric field structures.
The $\mathcal{S}$-map for e8-N/S is less prominent (Figure \ref{figure_field_dispersion}, upper right panel). 
Generally, B-field orientations appear to change 
less abruptly in the west (low values in $\mathcal{S}$) while the eastern half shows larger dispersion 
values. Similarly to e2-NW, $\mathcal{S}$ highlights the southern end of e8-S, where the B-field might
be locally dragged in, as an enhanced dispersion zone. Equally,  the northern end of e8-S, where the
curved cometary-like field structure is opening up, straightened and possibly pulled towards e8-N, 
also shows a patch of larger field dispersion.  Finally, the possibly converging streaming zone from southeast
to the height of e8-N
shows up as an elongated stretch with larger $\mathcal{S}$-values.
W51 North (Figure \ref{figure_field_dispersion}, bottom panel) shows an overall more uniformly small B-field dispersion. 
Two zones of enhanced dispersion
are identified symmetrically around N1, in its 
southeast and northwest.
This coincides with the pinching direction of the 
hourglass-like B-field. Two additional large-dispersion regions are seen slightly off the peaks of N3 and N4. 
In summary, in all three regions, W51 e2, e8 and North, the $\mathcal{S}$-parameter is often capturing areas that a visual inspection 
identifies as {\it magnetic field convergence zones} (Section \ref{result_magnetic_field}).

\begin{figure}
\includegraphics[width=9cm]{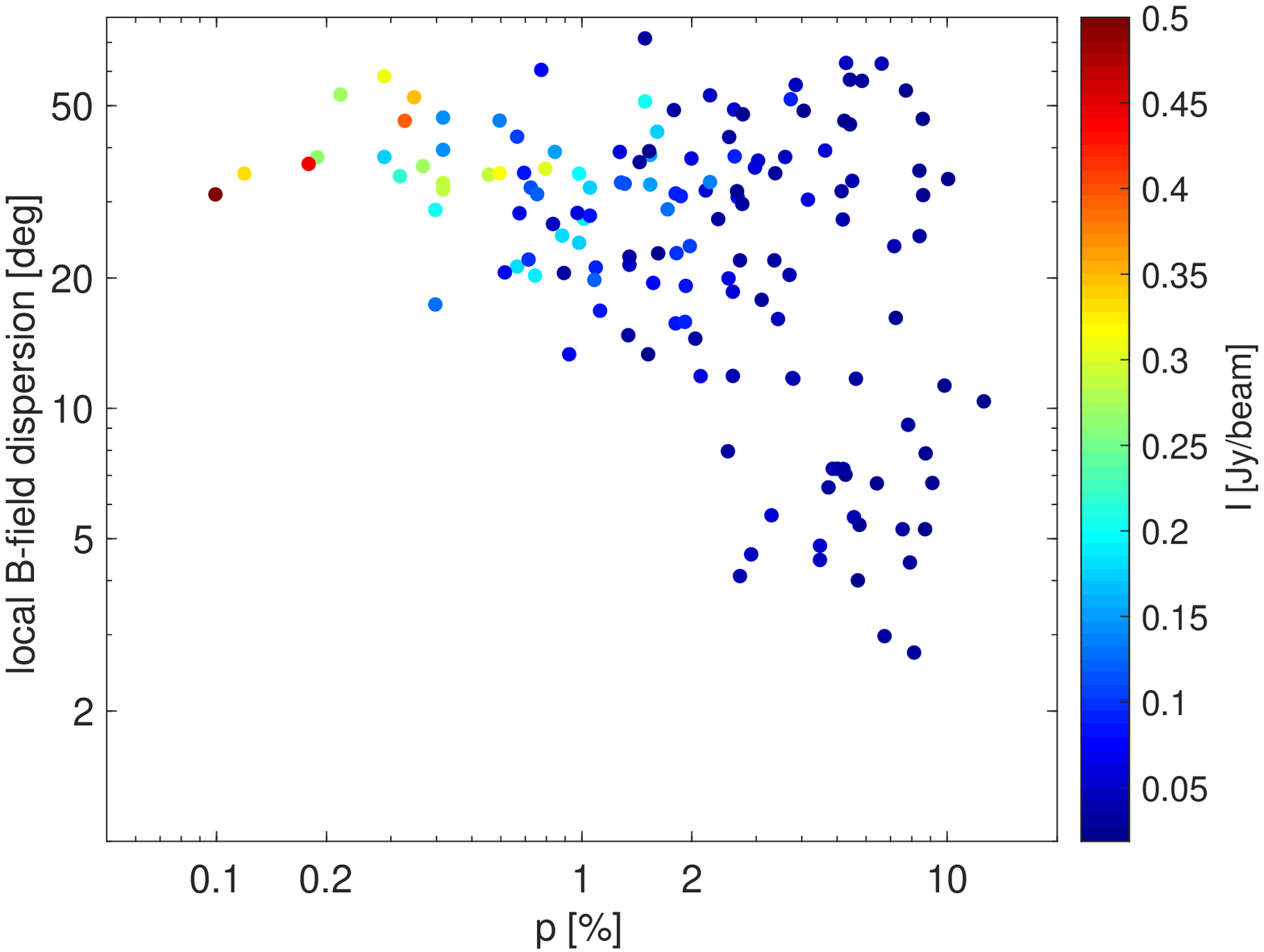}
\includegraphics[width=9cm]{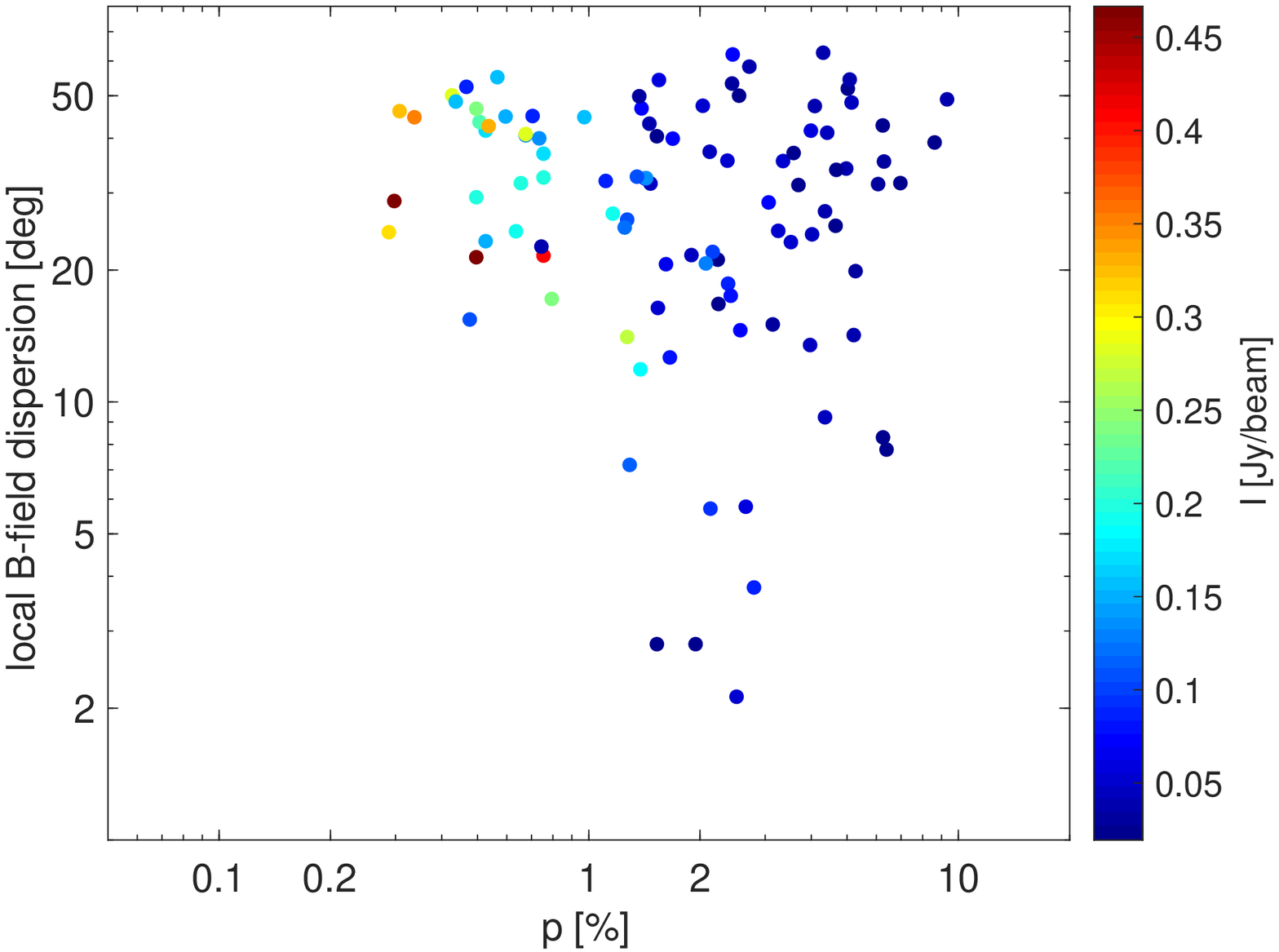}
\includegraphics[width=9cm]{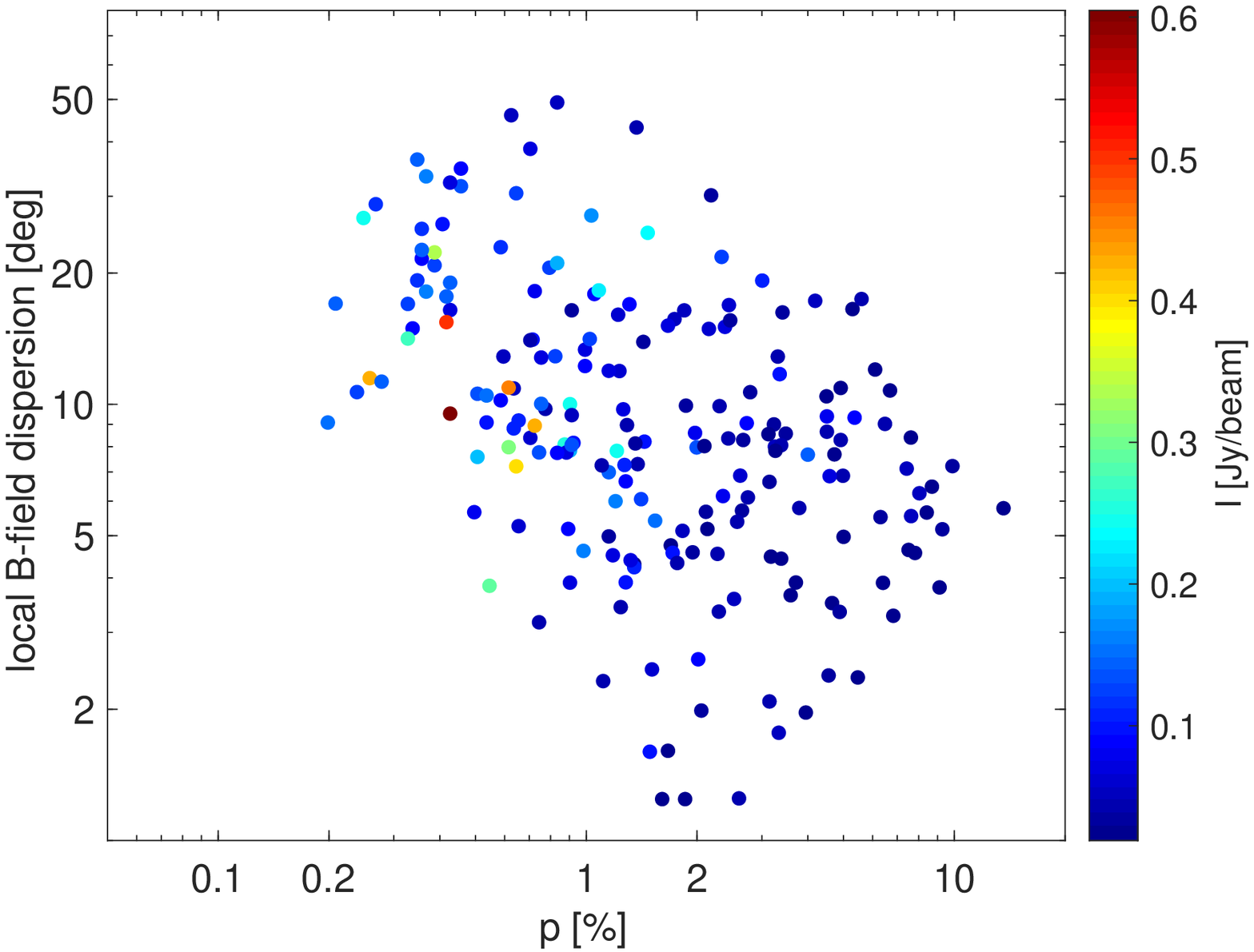}
\caption{Local magnetic field dispersion $\mathcal{S}$ ($r_{disp}=0\farcs2$) versus polarization percentage $p$ for e2 (top left), e8 (top right), and North (bottom). Stokes $I$ emission is color-coded.
Unlike the overgridded data displayed in Figure \ref{figure_field_dispersion}, data here are extracted from maps gridded to only half of the synthesized beam resolution.}
\label{figure_field_dispersion_2} 
\end{figure}

The spatial coincidence between lowest polarization fractions $p$ (Figure \ref{figure_pol_perc}) and 
largest dispersions $\mathcal{S}$ (Figure \ref{figure_field_dispersion}) is visible in many cores, 
e.g., stripe across e2-E and e2-NW, wings on N1, and peaks and offsets in N3 and N4. 
Figure \ref{figure_field_dispersion_2} shows the local B-field dispersion $\mathcal{S}$
as a function of polarization percentage $p$. While a large scatter in $\mathcal{S}$ is seen for peak polarizations, the 
smallest polarization percentages clearly converge towards the largest field dispersions. 
$\mathcal{S}$ appears to be anti-correlated with $p$, with a lower envelope (that traces the minimum polarization 
as a function of dispersion) and with a scatter that increases with $p$. 
Lowest polarization occurs at maximum dispersion values while these are seen across the entire Stokes $I$ range. 
$\mathcal{S}$ appears, thus, only weakly, if at all, correlated with $I$ (see Figure \ref{figure_all_correlations_polarizations} 
in Appendix). This is observed for all three regions, e2, e8, and North. 
The drop in polarization $p$ with increasing field dispersion can be interpreted as the cancellation of some
polarization signal due to more rapid changes in the field orientations. Hence, this might indicate that the 
observed field structures in W51 at a scale of 5~mpc ($\theta\sim 0\farcs26$) are not yet resolved
at those locations,
but underlying more rapidly changing
structures within our synthesized beam can be responsible for this anti-correlation. This same explanation holds 
for the observed anti-correlation between $p$ and Stokes $I$ (Section \ref{result_polarization} and Appendix Figure   
\ref{figure_all_correlations_polarizations}), assuming that $I$ is a fair tracer for the gas column density. 
Alternatively, an intrinsic lower grain alignment efficiency, due to varying densities and temperatures, might also
explain these two anti-correlations.
Our findings are in line with recent BLASTPol results for the Vela C molecular cloud \citep{fissel16} where 
a two-variable power-law empirical model is derived to describe the anti-correlation between $p$ and 
$\mathcal{S}$, and $p$ and column density $N$
on a scale of about 0.5~pc (observed at $500\mu$m with a resolution around $2\farcm5$).  
A decrease of $p$ with growing dispersion $\mathcal{S}$ on even larger scales is also noted in \citet{planckXX}. 
Our observations, thus, indicate a continuation of these anti-correlations on large parsec scales down to mpc scales.


Finally, we notice that the absolute polarization $I_p$ (Figure \ref{figure_I_pol}) and the local field dispersion $\mathcal{S}$ 
(Figure \ref{figure_field_dispersion}) 
-- although scattering in a broad band when comparing entire maps 
(Appendix, Figure \ref{figure_all_correlations_polarizations}) -- 
appear with similar structures in certain regions.  In particular, this is the case for e2-E and e2-NW, and e8 N/S
where maximum values in $\mathcal{S}$ always reflect minimum $I_p$ values.
The fact that no overall correlation is apparent, is likely because small and medium dispersions seem to come
with any values in $I_p$. This is especially the case for W51 North. 
While $\mathcal{S}$ vs $I_p$ seems to show a weaker less general correlation than $\mathcal{S}$ vs $p$
(Figure \ref{figure_field_dispersion_2}), the correlation tightens 
when limited to values around $I_{p,min}$ and $\mathcal{S}_{max}$.
As such, it namely exactly identifies the 
{\it magnetic field convergence zones} which appear with large $\mathcal{S}$, and both small polarization
percentage $p$ and small absolute polarization $I_p$.

\subsection{Gravity vs Magnetic Field: Local B-Field Resistance and Magnetic Channeling}

Section \ref{result_magnetic_field} has presented novel B-field features resolved with the ALMA 
0\farcs26 observations, namely (1) cometary B-field morphologies in e2-NW and e8-S, (2) convergence zones with  
symmetrical field structures in e.g., e2-E, and (3) possibly streamlined field morphologies between e.g., 
e8-S and e8-N, and north of e8-N towards e2. 
These new features are now starting to give the impression of actually seeing the dynamics of 
flowing material imprinted on and / or by the magnetic field morphologies.
Here, we are adding quantitative estimates that support this dynamical picture carved by
these detailed B-field morphologies.

{\it How important is the magnetic field in, e.g., e2-E, e2-W and e2-NW? In which cores can it
still slow down gravitational infall, where is the field already overwhelmed by gravity, and might
there be even local differences within the same core?}
We start our analysis from an ideal MHD force equation \citep[e.g.,][]{koch12a} that identifies the 
local direction of gravity through $\nabla\phi$ and the direction of the magnetic field tension force through $\mathbf{n}_B$. 
We impose 
the slight restriction that any change in the orthogonal field component is much smaller than
the total field strength, i.e. $\frac{\Delta B_{\bot}}{B}\ll 1$. 
This will hold for any spatially slowly changing
field functions\footnote{
On very small scales, this assumption might eventually fail for tangled magnetic 
fields if neighboring beams show large or abrupt changes in field orientations. 
There is, however, no indications of this to date from observed field morphologies.
The ALMA data presented here also still show smooth and continuous changes in almost
all locations. The local field dispersion $\mathcal{S}$ (Section \ref{discussion_polarization})
quantifies this with overall small values. 
}.
In return, this then allows us to simplify and combine the magnetic field hydrostatic
pressure and the field tension terms. With this, the force equation becomes \citep{koch12a}:
\begin{equation}
\rho\left(\frac{\partial}{\partial t}+\mathbf{v}\cdot\mathbf{\nabla}\right)\mathbf{v} =
-\mathbf{\nabla}P -\rho\nabla\phi+\frac{1}{4\pi}\frac{1}{R}B^2 \,\,\mathbf{n}_B,
                                                                                                             \label{mhd_momentum}
\end{equation}
where $\rho$ and $\mathbf{v}$ are the dust density and velocity, respectively.
$B$ is the magnetic field strength. $P$ is the hydrostatic dust pressure. 
$\phi$ is the gravitational potential resulting from the 
total mass contained inside the star forming region. $\mathbf{\nabla}$ denotes
the gradient. The field tension force (last term on the right hand side) 
with the field curvature $1/R$ is directed normal to the field line
along the unity vector $\mathbf{n}_B$.

\begin{figure}
\includegraphics[width=9cm]{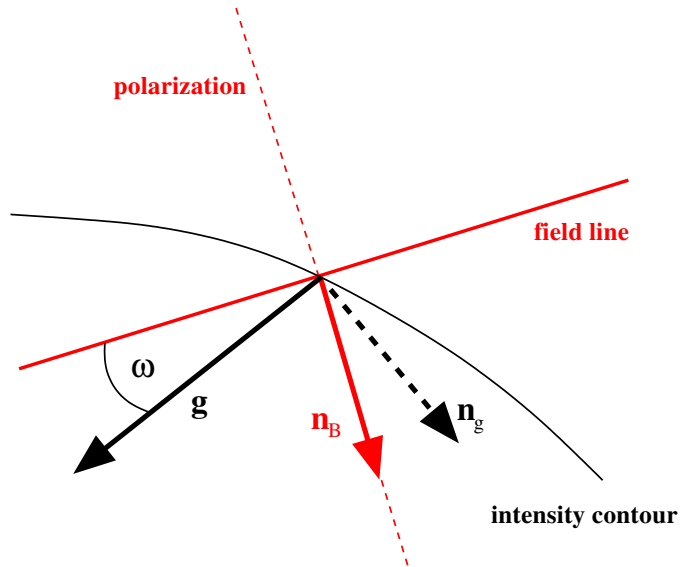}
\caption{Illustration of the angle $\omega$ that measures the deviation between the magnetic field orientation (in the case of submm dust polarization rotated by 90$^{\circ}$ from the originally detected polarization orientation) and the direction of local gravity $\mathbf{g}$. $\mathbf{n}_g$ is orthogonal to $\mathbf{g}$ and forms an orthonormal system with it.
$\mathbf{n}_B$ is the unity vector perpendicular to a measured B-field orientation along the direction of the field tension force.}
\label{figure_schematic_omega} 
\end{figure}

Since we are interested in comparing gravity and B-field, we rewrite the above equation
by projecting the direction of the field tension force onto the direction of gravity, 
i.e.,  $\mathbf{n}_B = \cos \omega\cdot \mathbf{n}_g + \sin\omega \cdot \mathbf{g}$, 
where $\mathbf{g}$ is a unity vector along the local direction of gravity $\nabla\phi$, $\mathbf{n}_g$ is the direction
normal to it forming an orthonormal system with $\mathbf{g}$, and $\omega$ is the angle between the 
local direction of gravity\footnote{
Calculating the local direction of gravity is introduced in \citet{koch12a}.
The observed dust emission {\it distribution} is assumed to represent the total mass {\it distribution} that generates 
the gravitational potential $\phi$. In order to measure the angle $\omega$, only the direction of the 
gravitational pull, $\nabla\phi$, is needed but not its magnitude. The total mass, linked to the dust emission through
an a priori unknown dust-to-gass mass ratio, is not needed and hence, it suffices to only consider the dust
distribution for this calculation.  The resulting local direction of gravity at a specific location is then derived by summing up 
all the surrounding pixelized dust emission weighted by $1/r^2$ along the direction to each pixel, where $r$ is the distance between that location and every surrounding pixel. 
In other words, every surrounding pixel is treated as a point mass that is exerting a gravitational pull on that specific
location.  The smallest scale that can be taken into account is given by the (synthesized) beam resolution. 
The largest size scale is defined by the largest distance to any detected emission.  We note that larger diffuser scales 
(emission) are filtered out in the  ALMA data but this is unlikely significantly affecting our results because (1) while 
already being weak, this emission is additionally less important due to a growing $1/r^2$ shielding; (2) the larger-scale 
emission often tends to be more symmetrically distributed, and resulting gravitational pulls can thus largely cancel
out. 
}
$\nabla\phi$ and the 
local B-field orientation,  
measured in a range between 0 and 90$^{\circ}$ (Figure \ref{figure_schematic_omega}).
Equation (\ref{mhd_momentum}) can then be rewritten as
\begin{equation}
\rho\left(\frac{\partial}{\partial t}+\mathbf{v}\cdot\mathbf{\nabla}\right)\mathbf{v} =
-\mathbf{\nabla}P -\rho|\nabla\phi|\mathbf{g}+\frac{1}{4\pi}\frac{1}{R}B^2 \,\sin\omega\,  \mathbf{g}
+ \frac{1}{4\pi}\frac{1}{R}B^2 \,\cos\omega \,\mathbf{n}_g.
                                                                                                             \label{mhd_momentum_2}
\end{equation}
In this way, the influence of the magnetic field along the direction of gravity $\mathbf{g}$, i.e.,  its effectiveness
opposing gravity, 
is quantified with the term $1/(4\pi R) B^2 \sin\omega$.  In particular, the factor $\sin\omega$
defines the fraction of $1/(4\pi R) B^2$ that can work against gravity to slow down or prohibit
infall, collapse, and any motion driven by gravity.
$\sin\omega$ is zero if the B-field is aligned with the local direction of gravity. 
In this case, independent of the field strength, the B-field can not resist gravity.  The magnetic 
field can maximally work against gravity if the field is orthogonal to gravity. Here, the precise 
effect will further depend on the field strength. 

\begin{figure}
\includegraphics[width=9.5cm]{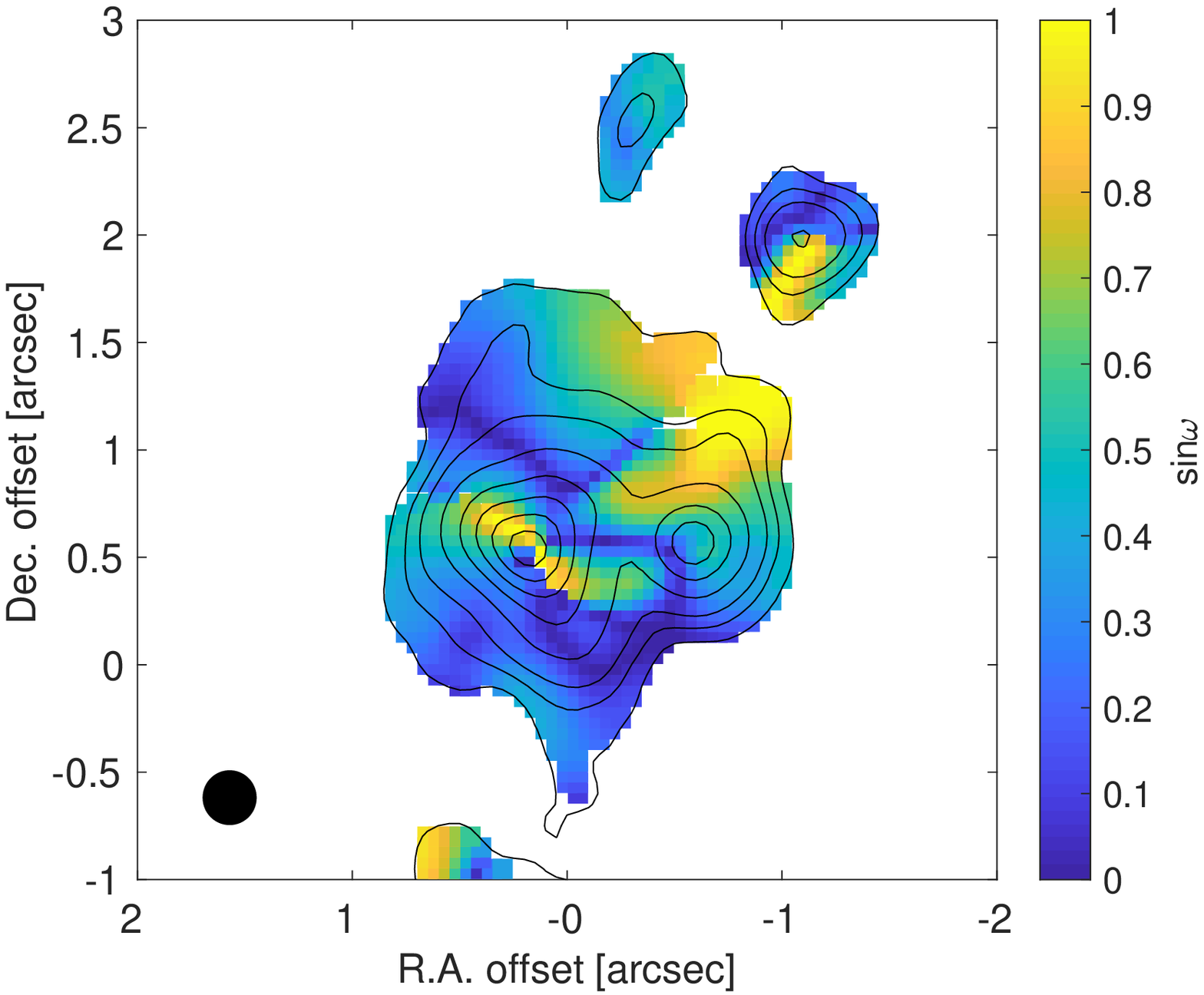}
\includegraphics[width=9.5cm]{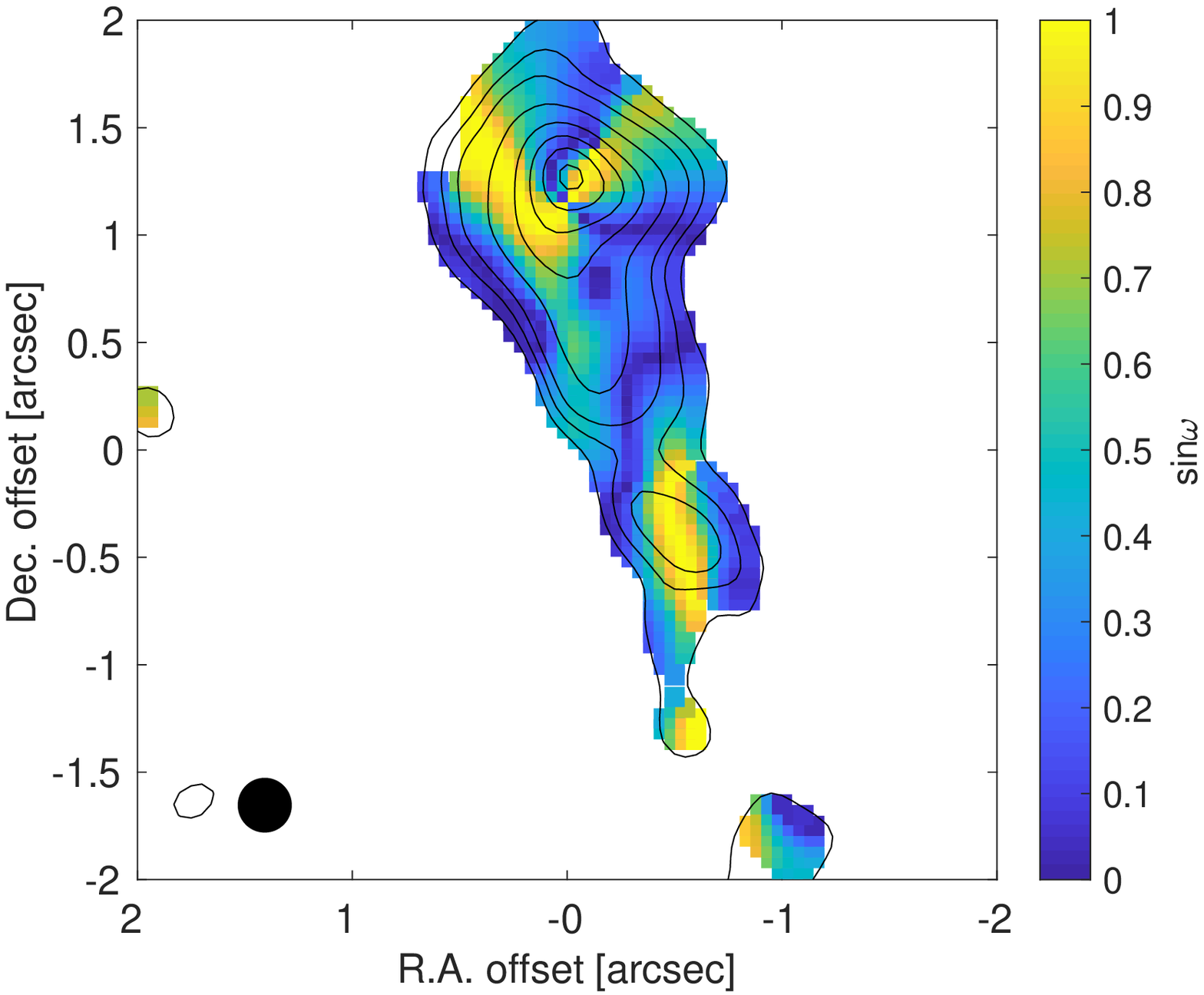}
\includegraphics[width=16cm]{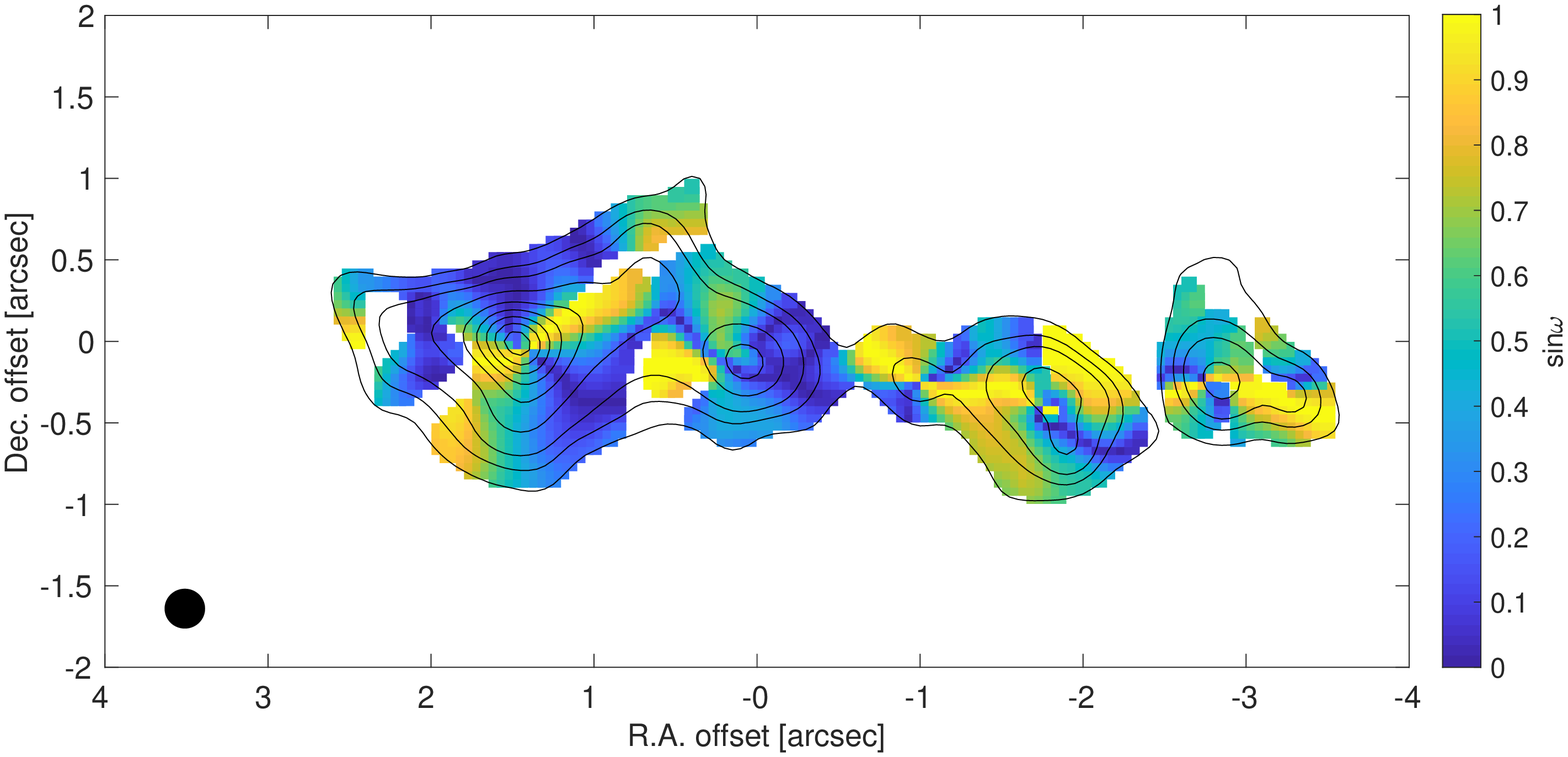}
\caption{$\sin\omega$-maps for e2 (top left), e8 (top right), and North (bottom).
$\sin\omega$ is in the range between 0 and 1, where $\sin\omega\sim 0$ and small values (blue)
indicate no or only minimal B-field effectiveness to oppose gravity, and $\sin\omega\sim 1$ (yellow, orange)
marks zones with maximum B-field effectiveness.  In the blueish zones, gravity can act most efficiently. 
Magnetic channels (dark blue, $\sin\omega\sim 0$) are within these zones.
Identical to the Figures \ref{figure_I_pol}, \ref{figure_pol_perc}, and \ref{figure_field_dispersion}, 
these maps are also overgridded for a sharper display of features. 
}
\label{figure_sin_omega} 
\end{figure}

Figure \ref{figure_sin_omega} shows $\sin\omega$-maps for the ALMA observations of W51 e2,  e8, and North.
It is evident that values are not random but appear in organized structures that can change with location.
$\sin\omega$ averaged over the entire W51 e2 map is small with 0.40 and a standard deviation of 0.27.
This indicates that e2, as an entity, is likely overwhelmed by gravity on the observed scales, with an overall
small magnetic field resistance.  e2-E displays some field resistance in the east that appears to grow towards
the center.  An additional zone of more significant B-field presence is in the north with $\sin\omega\sim 0.5$ or larger. 
Both zones occur at locations where the detected B-field orientations 
(panel (d) in Figure \ref{figure_composite_field}) are clearly more misaligned with the close-to-radial
gravity directions which result from close-to-circular emission contours of the individual cores. 
The area with largest B-field resistance ($\sin\omega\sim 1$) is around and north of e2-W.  In this 
zone, the B-field appears to be more tangential to the dust emission contours, possibly suggesting a pull towards 
e2-E and, therefore, a B-field opposing the local gravitational pull towards the center of e2-W.
The satellite core e2-NW displays a north-south asymmetry with minimal $\sin\omega$ values in the north 
and values close to one in the south. This is opposite to the observed north-south symmetry in the cometary 
B-field morphology, and in both $I_p$ and $p$ (Figures \ref{figure_composite_field}, \ref{figure_I_pol}, \ref{figure_pol_perc}). 
What likely drives this asymmetry is the massive e2-E/e2-W complex that is pulling e2-NW
towards south. As a consequence, local gravity directions in the e2-NW core deviate from being simply radial towards its
emission peak. This effect is most significant in the south, thus, leading to large misalignments $\omega$ between 
field orientations and gravity.  

W51 e8 shows a very similar overall B-field effectiveness to oppose gravity with an average $\sin\omega$ value of 0.41
and a standard deviation of 0.29.  e8-N dominantly reveals small values around 0.2-0.5, as expected from the 
likely pulled-in field morphology in these locations (panel (e) in Figure \ref{figure_composite_field}).
The exception is the eastern side with values up to one. 
These highest values coincide with field segments that are more tangential to contours and the possible bending away 
from e8-N (Section \ref{result_magnetic_field}). Except for a western triangular section with $\sin\omega\sim 0.7$, the elongated bridge between e8-N and e8-S mostly displays small values 
around 0.1-0.3. This is in agreement with the visual impression of a streamlined B-field morphology 
that is probably driven by the locally
dominating gravitational center e8-N.
The east-west symmetry in the possibly cometary B-field morphology in e8-S is not completely preserved in $\sin\omega$. 
This is likely because the dominating mass, e8-N, is located off the north-south axis. 
Overall, e8-S is close to maximally resisting gravity in its center (with $\sin\omega\sim 1$).  
At its western and eastern sides, local gravity is aligned with the B-field, hinting that
e8-S can accrete material from the two sides.
We note that, although showing maximum B-field 
effectiveness in the center, this does not yet mean that the B-field is dominating over gravity. 

W51 North shows varying $\sin\omega$ structures in every core, with an overall average of 0.47 and a standard
deviation of 0.30.
The most massive eastern core (N1) displays small 
values around $\sim$0.1 to 0.2 in a fan-like opening along a northeast-southwest axis. This area precisely overlaps
with the polar regions of the possible hourglass field structure in this core (panel (h) in Figure \ref{figure_composite_field}). 
The very small values in $\sin\omega$
indeed suggest, as expected, that gravitational infall / collapse can easily proceed along this direction, and that
the B-field is here mostly only channeling material.  In contrast to that, 
along the northwest-southeast axis -- where the B-field is
more pulled in -- $\sin\omega$ is reaching maximum values close to one, indicating maximum field resistance.
The next massive core to the west (N2, centered around R.A. offset $0\arcsec$) reveals another fan-like low $\sin\omega$
area in the west with gradually growing values at its eastern end. As outlined in Section \ref{result_magnetic_field}
this can explain a scenario where infall / collapse can occur locally in the western zone while the eastern end starts
to feel the gravitational pull by the more massive core N1 to the east, leading to a gradual bending of the field lines 
away from N2. 
This gradual bending then leads to a transition zone in between N1 and N2 where gravity and field segments are 
misaligned (hence, large $\sin\omega$ values), before they are aligned again in the gravitationally dominated zones
in N1. 
The remaining two cores in the west around R.A. offset $-2\arcsec$ (N3) and R.A. offset $-3\arcsec$ (N4) display more
complicated and finer structures. A possible feature is that small $\sin\omega$ values are typically found in the 
upper (northern) and lower (southern) mid-planes of the cores, while most maximum values appear along an 
east-west axis. This reflects that many field segments show a prevailing north-south orientation, which makes them
being closely aligned with gravity in many places. The exceptions are the mid-planes, in particular the eastern and 
western ends of
N3, the western end of N4,  and the  connecting bridge between N3 and N2 where the B-field experiences a competition between gravitational pull
towards N3 and pull towards N2.  

Note that {\it $\sin\omega$ is different from the magnetic field-to-gravity force ratio, $\Sigma_B$}, in our earlier 
analyses \citep{koch12a,koch12b,koch13,koch14}. $\Sigma_B$ measures the local ratio between magnetic field force
and gravity (in a range between zero to larger than one) by {\it solving} Equation (\ref{mhd_momentum}),
identifying the local direction of gravity and the field tension in an observed map.
It, thus, compares and quantifies, in an absolute sense, the relative importance between magnetic field and gravity.
Solving Equation (\ref{mhd_momentum}) is based on the additional assumption of identifying an observed emission gradient
direction with the inertial term in the MHD force equation.  Appropriateness and possible uncertainties
of this assumption are discussed in detail in \citet{koch13}.
$\sin\omega$ is not relying on solving Equation (\ref{mhd_momentum}). It merely projects the field force onto 
the local gravity direction, and it is, thus, free of the above assumption.
Its shortcoming is that it can only capture the zones where gravity is clearly dominating over the B-field,
i.e., where the B-field is dynamically unimportant ($\sin\omega \sim 0$ or small). For larger values, $\sin\omega\sim 1$,
the absolute field strength becomes relevant for a quantitative absolute comparison against gravity.
A detailed comparison between $\sin\omega$ and $\Sigma_B$, together with maps for the field strength $B$, 
will be presented in a forthcoming work.  An initial comparison is showing a close structural resemblance between
$\sin\omega$- and $\Sigma_B$-maps. Since projection effects cancel out or are minimal for $\Sigma_B$ --
because $\Sigma_B$ is the ratio of two forces with each force direction being subject to the same or similar 
inclination angle \citep{koch12a} -- this close resemblance argues for $\sin\omega$ being able to distinguish between 
zones of minimal and maximal B-field effectiveness without any significant bias due to unknown projection effects. 

On a final speculative note, we stress that $\sin\omega$ is clearly not random. Moreover, within zones of low 
field effectiveness (small $\sin\omega$ values), there are channels with $\sin\omega\sim0$ (Figure \ref{figure_sin_omega}). 
Many of these channels appear in magnetic field convergence zones. 
In these {\it magnetic channels}, gravity can act unimpededly. If this is the case, this would indicate that along certain
directions, infall and collapse can proceed in free-fall time while in other zones they are significantly slowed down or 
even completely brought to a halt.  This interpretation is, however,  at or already beyond the limit of the current resolution. 
Whether this speculative scenario indeed is correct, needs to be further probed with even higher-resolution data. 
If correct, the existence of a network of magnetic channels might have an interesting implication for the star formation
rate.  Hypothesizing a channel width of $\sim0\farcs15$ (about half of our synthesized beam leading to a marginal
detection), one such channel from the rim to the center would comprise about 0.4\% of the entire volume of a 2\arcsec-
diameter sphere as, e.g., in W51 e2. A network of 10 channels would then reduce the star formation rate to 4\%, assuming that the entire mass inside the channels ($\sin\omega \sim 0$) is converted into stars, and that all the material outside
($\sin\omega$ large) is held back by the B-field. 
Checking observed star formation rates against future high-resolution B-field structures might hence also provide a test
to further probe the $\sin\omega$ tool.

%
%

\subsection{Comparing to Larger Scales: B-field and Polarization Properties in the pc-Scale Envelope}
                                                                                               \label{discsussion_larger_scale}

While the high-resolution ALMA data are making the small-scale dynamics visible in the B-field
morphologies,  a remaining key question is how the larger W51 e2, and the more elongated and filamentary 
e8 and W51 North are formed in a first place.  To that purpose, we are here additionally comparing
unpublished
SMA data on larger scales that also detect the polarized emission on the bridge
between e2 and e8, and in the more outer peripheral regions towards e2 and e8 (Figures \ref{figure_composite_pol},
\ref{figure_composite_field}).
A detailed analysis of the dynamics along the e2/e8 bridge is presented in 
\citet{koch17b}.

\begin{figure}
\includegraphics[width=9cm]{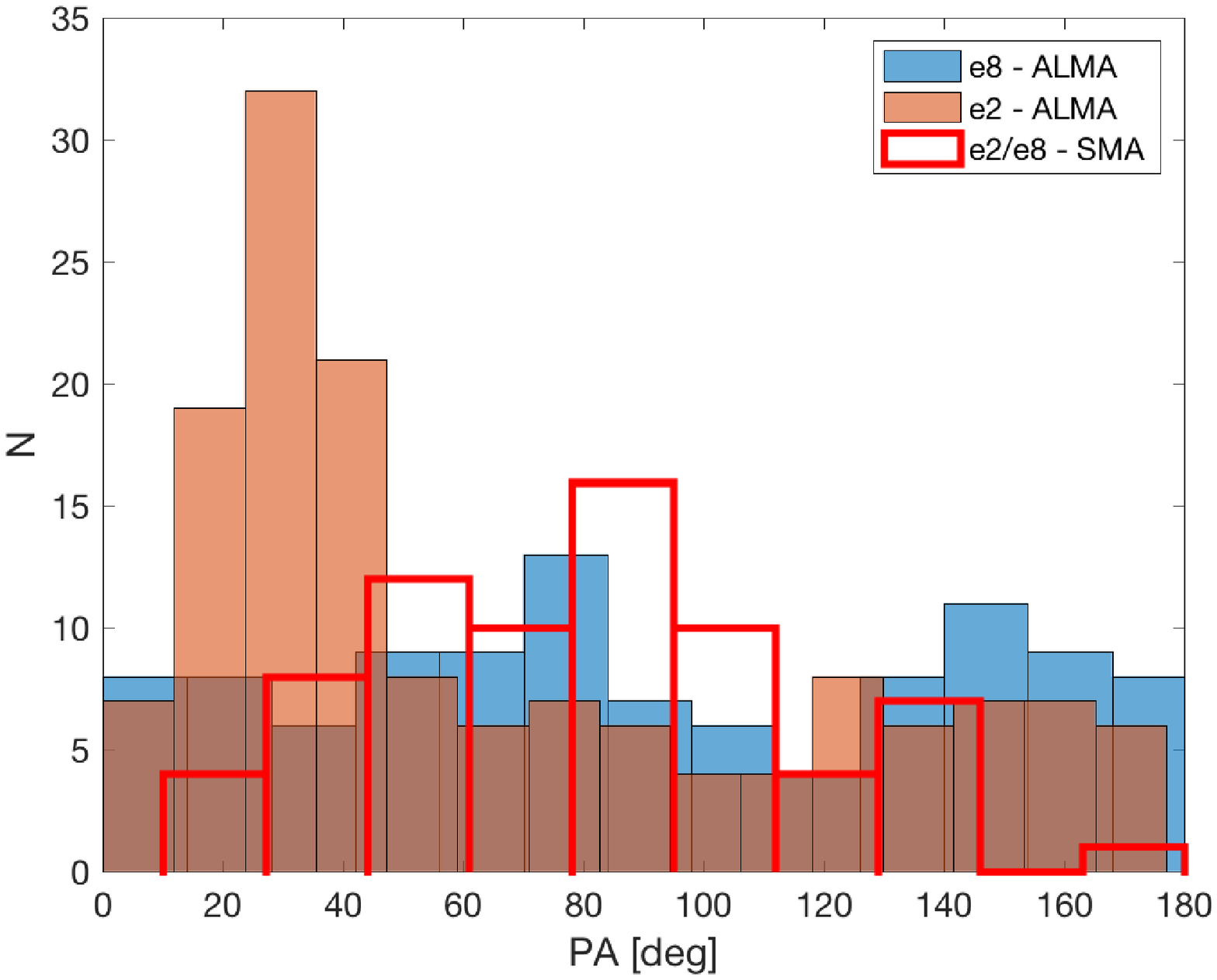}
\includegraphics[width=9cm]{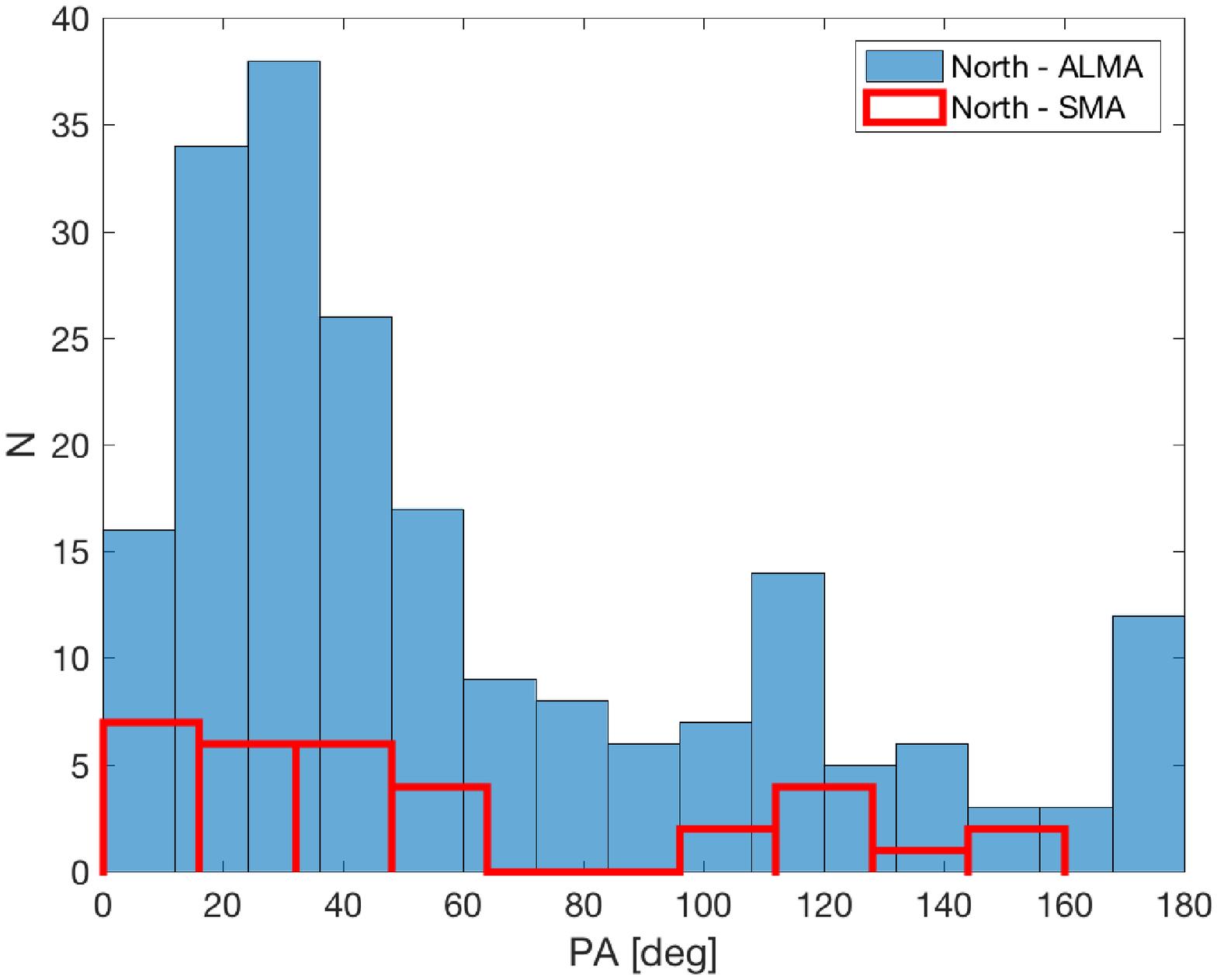}
\caption{Left panel: Histograms of B-field orientations for the SMA combined e2/e8 ($\theta\sim2\arcsec$, panel (a) in Figure \ref{figure_composite_field}), and the separate
$\theta\sim0\farcs26$ ALMA e2 and e8 (panels (d) and (e) in Figure \ref{figure_composite_field}).
Right panel: identical but for SMA North and ALMA North.}
\label{figure_hist_e2_e8_North} 
\end{figure}

The W51 e2/e8 complex is connected with a bridge where the B-field segments appear to be
bending away from e8 and gradually more directed towards e2 (panel (a) in Figure \ref{figure_composite_field}). 
Beyond a distance of about one beam size from the e2 and e8 peaks, the majority of the B-field
segments starts to display a prevailing east-west orientation. The outer peripheral zones, thus, clearly
reveal field orientations that are perpendicular to the north-south axis of e2/e8, likely probing 
accretion scales that are very different from the inner much denser regions where the field structures
likely start to be shaped by gravity. 
The histogram of B-field orientations - capturing the larger envelope on a $\sim0.5$~pc scale -- reflects this with a peak around 90$^{\circ}$, i.e., east-west orientation (Figure \ref{figure_hist_e2_e8_North}, left panel). Histograms for the 
resolved cores at a $\sim5$~mpc scale spread out over the full range in PA from 0 to 180$^{\circ}$, indicating a change
from a single prevailing orientation to a more uniform and broader distribution that stands for a more azimuthally symmetrical 
field configuration. 
While e8 is fairly uniform, e2 additionally shows a peak around $\sim30^{\circ}$ which coincides with the 
orientation 
of magnetic field and core major axis
already imprinted on the larger scale (dotted square in panel (a) in Figure \ref{figure_composite_field}),
about $\sim30^{\circ}$ off the north-south axis.

W51 North shows a similar scenario. The SMA observations with a 
coarser resolution around 2\arcsec (panel (f) in Figure \ref{figure_composite_field}) show 
one dominating core in an extended envelope along an 
east-west direction.  At the R.A of the core, field segments are north-south aligned, perpendicular 
to the source's longer axis. Further to the west, B-field orientations are gradually bending, pointing
towards the main peak.  
This is seen in the histogram with two separate groups around $\sim 0$ to $40^{\circ}$ and around 
$\sim120^{\circ}$ (Figure \ref{figure_hist_e2_e8_North}, right panel).
The ALMA data (panel (h) in Figure \ref{figure_composite_field}), resolving the individual cores on a $\sim5$~mpc scale, again show orientations
spreading over the full PA range. Their distribution reflects again azimuthally symmetrical field configurations, with 
one peak around $\sim 30^{\circ}$ which results from the main core that appears to be rotated by $\sim 30^{\circ}$
with respect to the larger-scale north-south orientation. 

In conclusion, for both e2/e8 and North,  denser cores (e2, e8, and a chain of cores in North) appear to form along 
an axis perpendicular to the larger envelope-scale B-field.  Their most massive cores (e2 and N1) have their B-fields
oriented close to the envelope-scale field, with offsets around $\sim30^{\circ}$. 
However, when further zooming in with higher resolution, 
B-field configurations generally are much more azimuthally symmetrical with further substructures that are likely 
decoupled from the larger-scale B-field and governed by their own dynamics. 
In general, the field configurations seen with the SMA on envelope scales are suggestive for a scenario where the B-field is channelling material from east and west (W51 e2/e8) and from north and south (W51 North, \citet{tang13}) to a mid-plane. 
The higher resolution SMA and ALMA data, indeed, confirm the locations of the denser cores to be aligned
along a north-south (W51 e2/e8) and east-west (W51 North) axis.

\begin{figure}
\includegraphics[width=18cm]{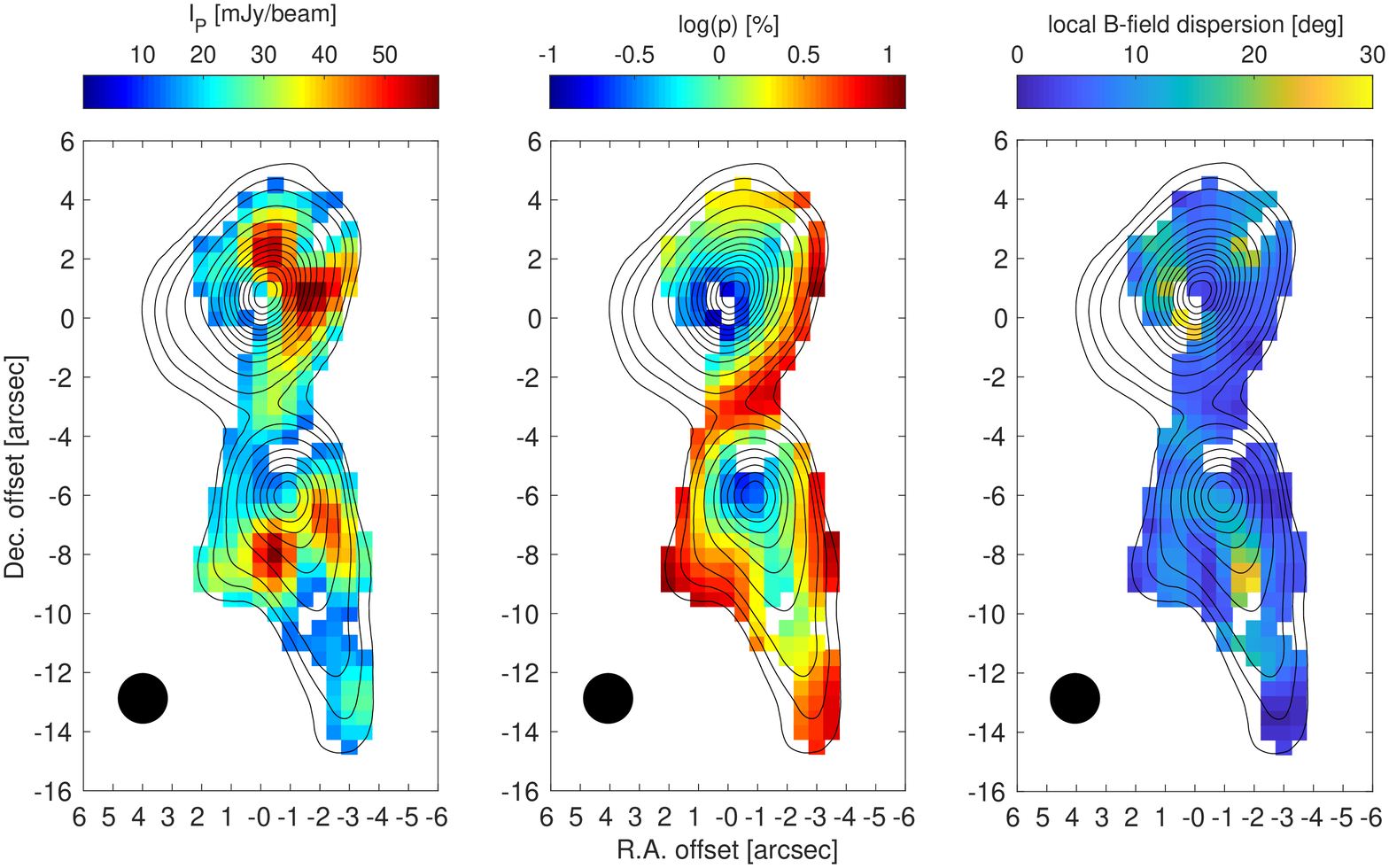}
\includegraphics[width=18cm]{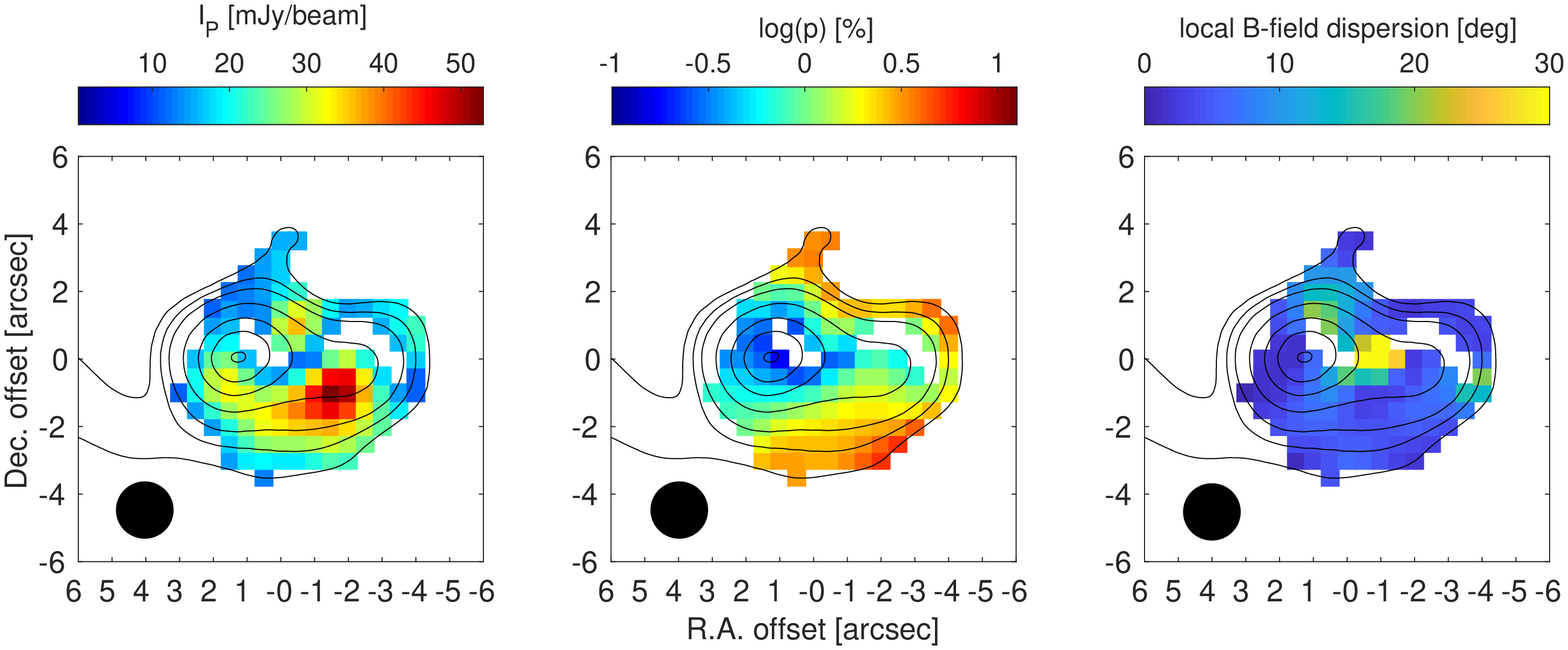}
\caption{\small Absolute polarization $I_p$, polarization percentage $p=I_p/I$, and local B-field dispersion $\mathcal{S}$
for the SMA subcompact array observations of W51 e2/e8 (upper panels) and North (lower panels) with a resolution
$\theta\sim2\arcsec$. The displayed data are overgridded to $0\farcs5$.
Contours are dust continuum with levels as described in Figure \ref{figure_composite_pol},
}
\label{figure_sma_subc_comb_1} 
\end{figure}

\begin{figure}
\includegraphics[width=9cm]{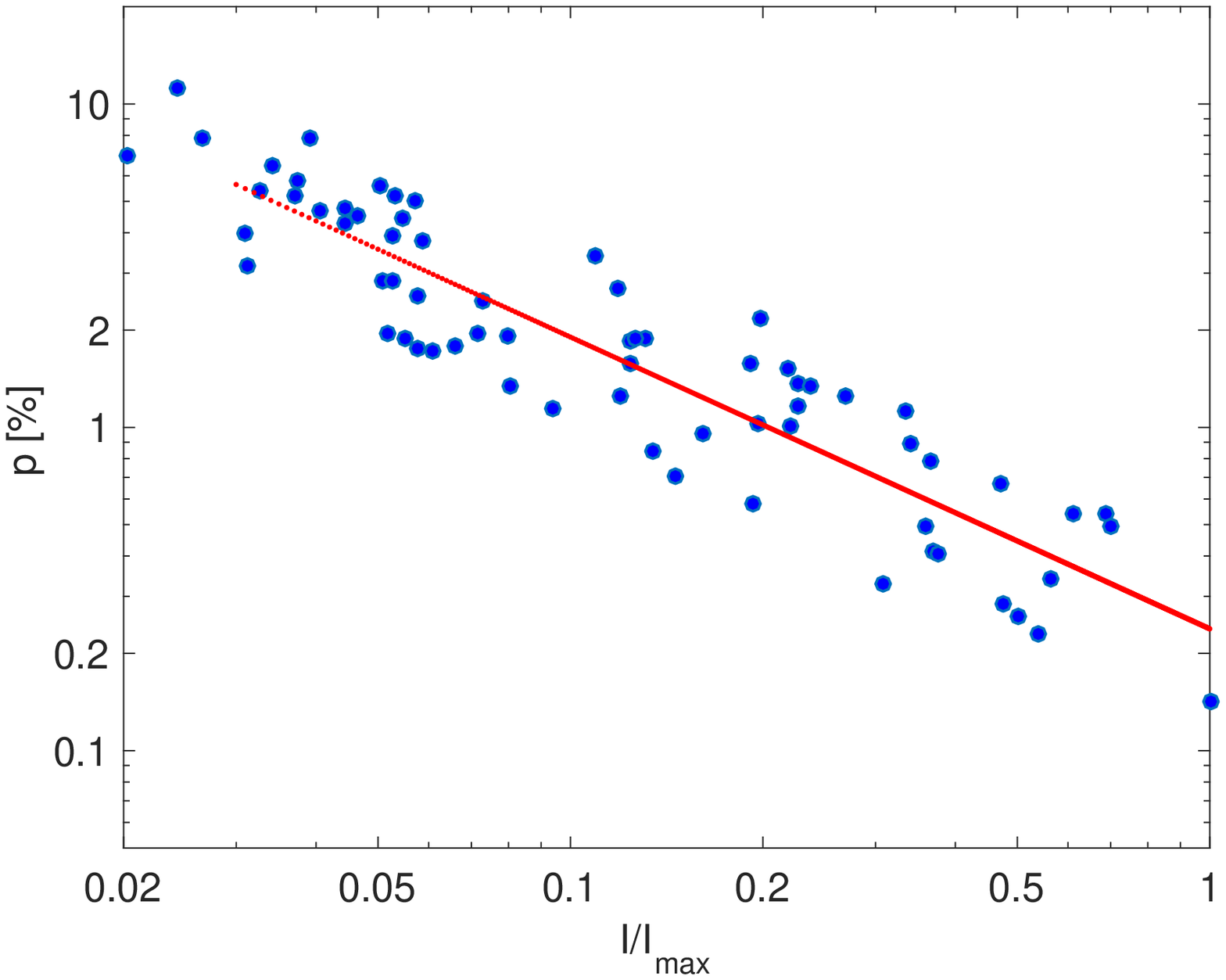}
\includegraphics[width=9cm]{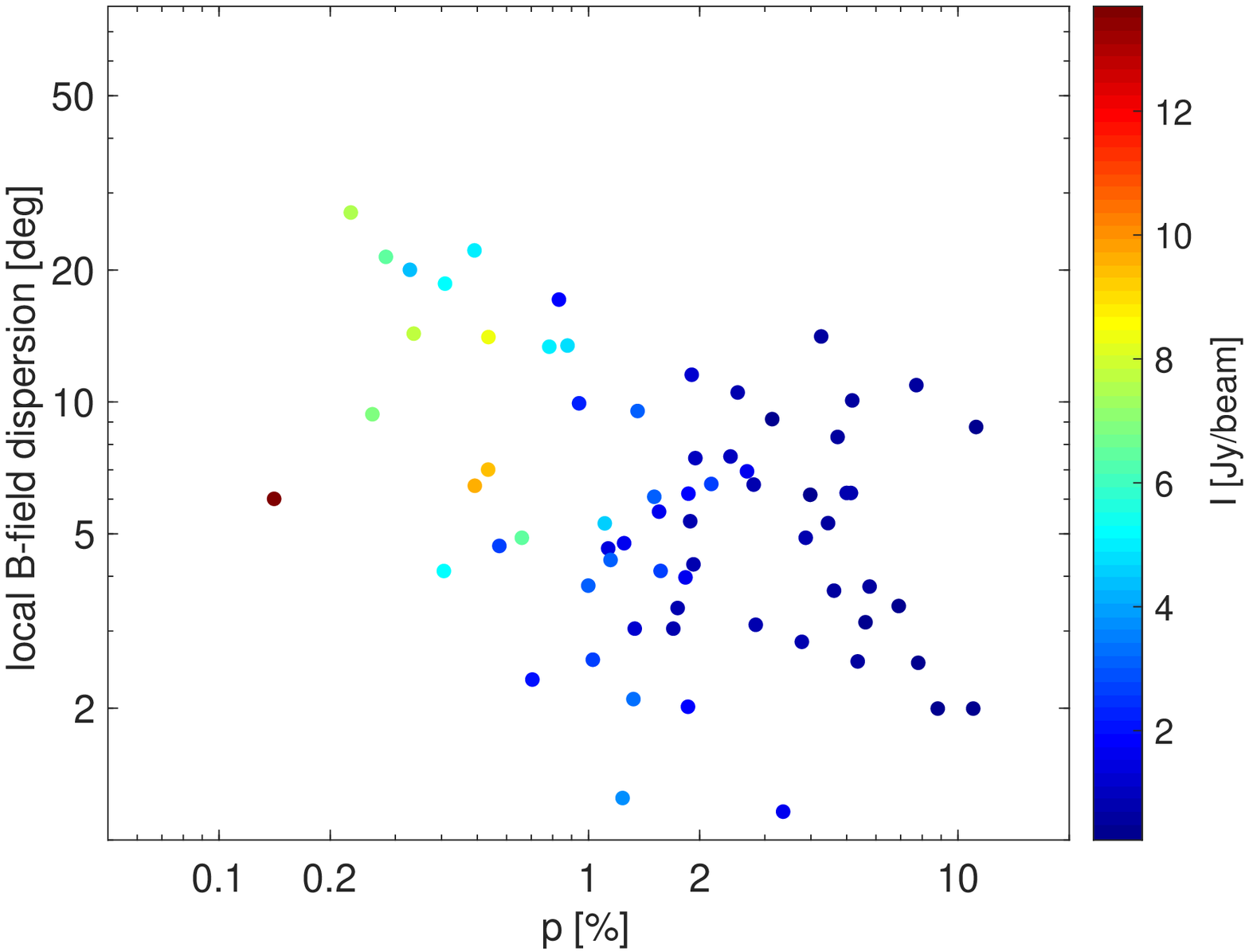}
\includegraphics[width=9cm]{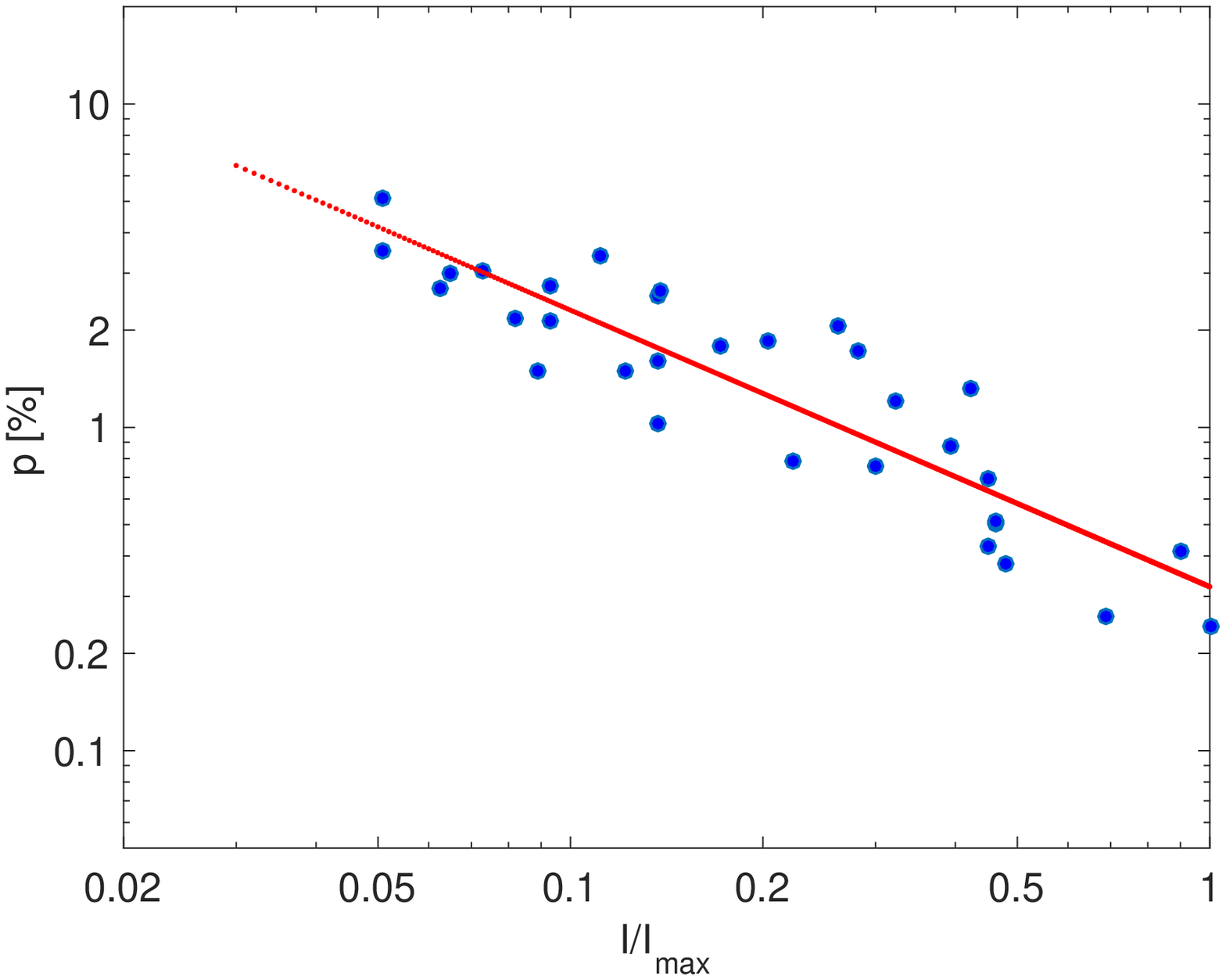}
\includegraphics[width=9cm]{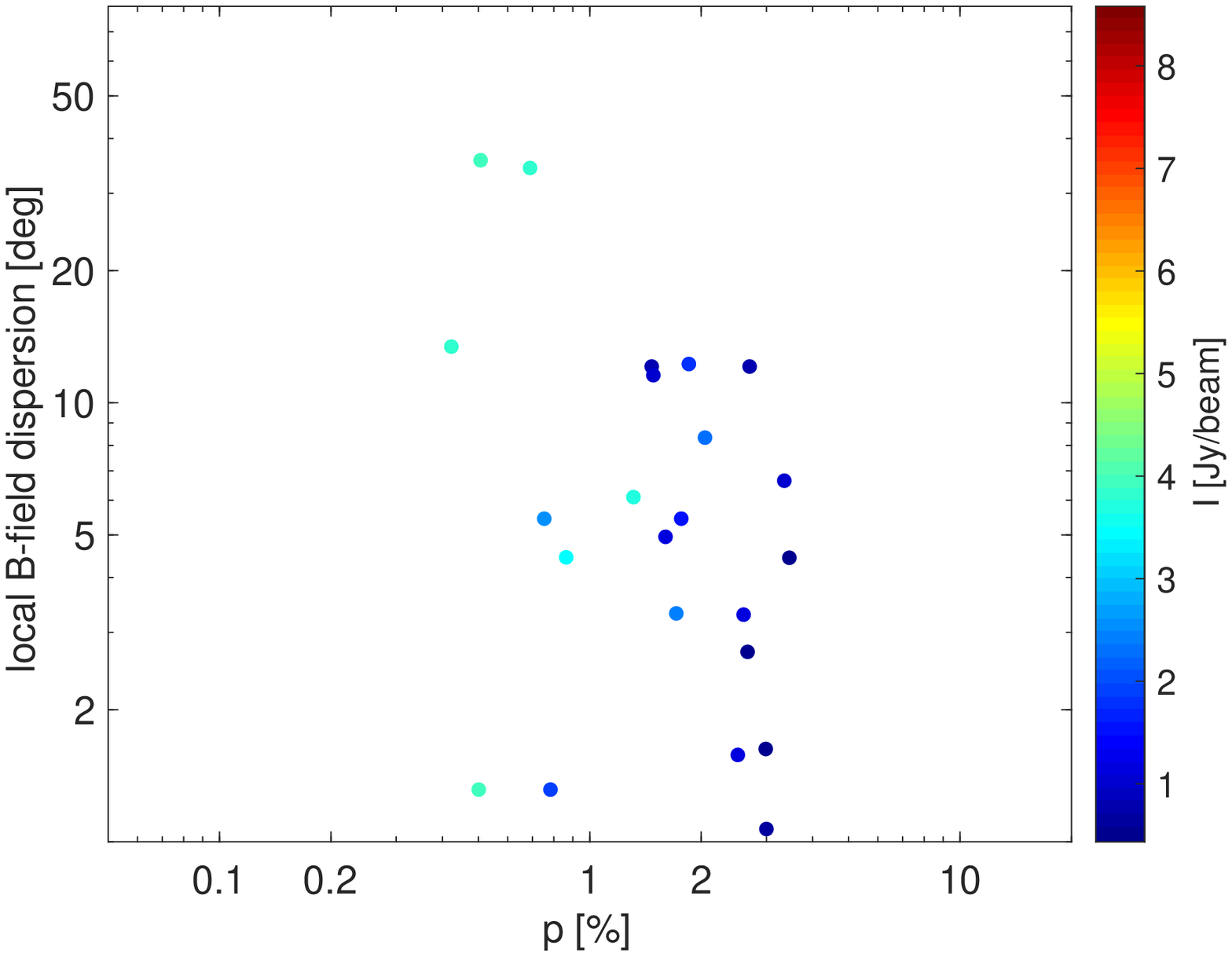}
\caption{Polarization percentage $p=I_p/I$ versus Stokes $I$, normalized to $I_{max}$ (left panels) and 
local B-field dispersion $\mathcal{S}$ versus $p$ (right panels) for W51 e2/e8 (upper panels) and North (lower panels)
for the SMA subcompact array observations in the panels (a) and (f) in Figure \ref{figure_composite_pol} 
and \ref{figure_composite_field}.
Unlike the overgridded data displayed in Figure \ref{figure_sma_subc_comb_1}, the data here are extracted from maps 
gridded to only half of the 
synthesized beam resolution (Figures \ref{figure_composite_pol} and \ref{figure_composite_field}).
The red solid lines are the best-fit power laws with indices -0.90 (e2/e8) and -0.86 (North).}
\label{figure_sma_subc_comb_2} 
\end{figure}

An analysis analog to Section \ref{discussion_polarization} for the SMA envelope-scale data shows similar trends.
The local B-field dispersion $\mathcal{S}$ (Figure \ref{figure_sma_subc_comb_1}) 
is picking up the locations where the field is changing from 
a rather uniform orientation (east-west in e2/e8 and north-south in North) to gravitational pull-in
(northern and southern end in e2, southern end in e8, and western end in North).  Features in absolute
polarization $I_p$ are also seen at this resolution with two 
maxima
in both e2 and e8 off their Stokes $I$
peak positions, and with one 
maximum
in North, west of its peak (Figure \ref{figure_sma_subc_comb_1}). 
Relative polarization percentages $p=I_p/I$ are -- similar to the ALMA data -- up to about 10\% with maximum
values at the lowest Stokes $I$ contours.  This confirms the $p$ vs $I$ anti-correlation also on the 
$\sim0.5$~pc envelope scale (Figure \ref{figure_sma_subc_comb_2}, left panels).
The only noticeable difference in this comparison with ALMA is the much more scattered
$\mathcal{S}$ vs $p$ relation, indicating no correlation between these two observables on this scale in these sources
(Figure \ref{figure_sma_subc_comb_2}, right panels).

\section{Summary and Conclusion} \label{summary}

We are presenting the first ALMA continuum polarization observations towards the high-mass star-forming regions 
W51 e2, e8, 
and North in Band 6 (230~GHz) with a resolution of $0\farcs26$. We further propose a diagnostic -- the angle $\omega$
between the local magnetic (B) field orientation and the local gravity direction -- as a way to assess the effectiveness of 
the B-field to oppose gravitational pull.
Our main results are summarized in the following.

\begin{enumerate}

\item{\it Polarization Structures.}
Polarized emission is clearly detected with ALMA across all three sources W51 e2, e8, and North. Polarization holes in the 
earlier SMA data are resolved, implying that such holes and depolarization zones can be signposts for more 
detailed or finer underlying B-field structures. The absolute polarized emission $I_p$ shows complicated structures 
that do not simply correlate or anti-correlate with total Stokes $I$.  Polarization percentages $p$ are measured 
from around 0.1\% to 10\%.  While all three sources reveal an anti-correlation between $p$ and $I$ with slopes
around $-1$, this anti-correlation shows a large scatter of almost an order of magnitude. This suggests
additional physics that is not yet captured in this simple anti-correlation.
SMA observations with $\theta\sim2\arcsec$ also show a $p$ vs $I$ anti-correlation with similar slope and range in polarization percentage, but with a possibly smaller scatter. 

\item{\it Magnetic Field Morphologies.}
The B-field structures in all three regions are organized, coherent and connected. Additionally, detailed new substructures
are resolved, revealing cometary B-field morphologies in satellite cores, convergence zones with symmetrical B-field
structures, and possibly streamlined B-field morphologies. Many of the cores display structures that resemble 
gravitationally bent or pulled-in field lines at one end of the core, while field lines at the other end of the core appear
to be dragged away towards the neighbouring next more massive core. This might support a scenario where local collapse can 
start in a small core while this core as an entity is dragged towards the next larger gravitational center. 
The larger envelope-scale B-field morphologies, captured with the SMA $\theta\sim2\arcsec$ observations, reveal 
prevailing field orientations perpendicular to the direction of the aligned higher-resolution ALMA cores. The 
bridge between W51 e2 and e8 reveals field lines that are gradually more bent towards the more massive e2.

\item{\it Magnetic Field Dispersion and Convergence Zones.}
Similar to larger-scale observations by {\it Planck} and BLASTPol, a connection between the local B-field 
dispersion $\mathcal{S}$ -- capturing by how much the local field orientation varies with respect to its surrounding --
and polarization percentage is also seen in the high-resolution ALMA data. In particular, 
a close spatial coincidence between lowest polarization percentages $p$ and largest dispersions $\mathcal{S}$
is evident in 
magnetic field convergence zones, where symmetrical B-field structures appear to converge from two sides.
These convergence zones always also have smallest absolute polarized emission $I_p$. 
The drop in both $I_p$ and $p$ towards zones of growing $\mathcal{S}$ can be interpreted as the cancellation
of polarization signal due to rapidly changing field structures that are still not resolved. 

\item{\it Local Magnetic Field Effectiveness Opposing Gravity.}
The direction of the local B-field tension force can be projected onto the local direction of gravity by means
of the measurable angle $\omega$ between a B-field orientation and local gravity. In this way, $\sin\omega$
measures the fraction of the field tension force that can work against gravity.
Maps of $\sin\omega$
for all three sources W51 e2, e8, and North are not random but present organized patterns. 
Zones where $\sin\omega$ is small (gravity and B-field nearly aligned) 
indicate that the B-field is very ineffective in slowing down gravity. 
Any motion driven by gravity can proceed with little or almost no obstruction from the B-field.
For regions with larger $\sin\omega$ values the absolute field strength and gravity need to be known to quantify
the role of the B-field.  Narrow sectors with $\sin\omega\sim0$ lead to the speculation of magnetic channeling where
infall and collapse could proceed in free-fall time.

\end{enumerate}

\begin{table}[]
\centering
\begin{tabular}{|l|cccc|}\hline
Name  & R.A. & Decl. &  Size & Flux\\
        & (J2000) & (J2000) & (")$\times$(")  &  (Jy) \\\hline
        \hline
W51 e2-E & 19:23:43.96 & 14:30:34.56 &  0.64$\times$0.54 & 2.54\\ 
W51 e2-N & 19:23:43.93 & 14:30:36.53  & 0.52$\times$0.11  & 0.10  \\ 
W51 e2-W & 19:23:43.91 & 14:30:34.59 & 0.44$\times$0.39 & 1.04 \\
W51 e2-NW & 19:23:43.88 & 14:30:35.98 & 0.28$\times$0.22 & 0.22 \\
W51 e8-N & 19:23:43.90 & 14:30:28.14 & 0.81$\times$0.51 & 2.65 \\
W51 e8-S & 19:23:43.87 & 14:30:26.63 & 0.67$\times$0.31 & 0.34 \\
W51N1 & 19:23:40.05 & 14:31:5.47 & 0.73$\times$0.51 & 2.88 \\
W51N SMA2-E & 19:23:40.00 & 14:31:5.73 & 0.88$\times$0.73 & 1.49 \\
W51N2 & 19:23:39.96 & 14:31:5.41 & 0.62$\times$0.41 & 0.96 \\
W51N3 & 19:23:39.83 & 14:41:5.11 & 0.74$\times$0.52 & 0.89 \\
W51N4 & 19:23:39.75 & 14:41:5.26 & 0.68$\times$0.45 & 0.59 \\
\hline
\end{tabular}
\caption{Cores in W51 e2, e8, and North, identified from dust continuum in Figure \ref{figure_composite_pol}.
Positions, deconvolved sizes, and integrated flux densities 
are all measured at 230~GHz in Band 6 with ALMA .}
\label{table_cores}
\end{table}

\begin{figure}
\includegraphics[width=18cm]{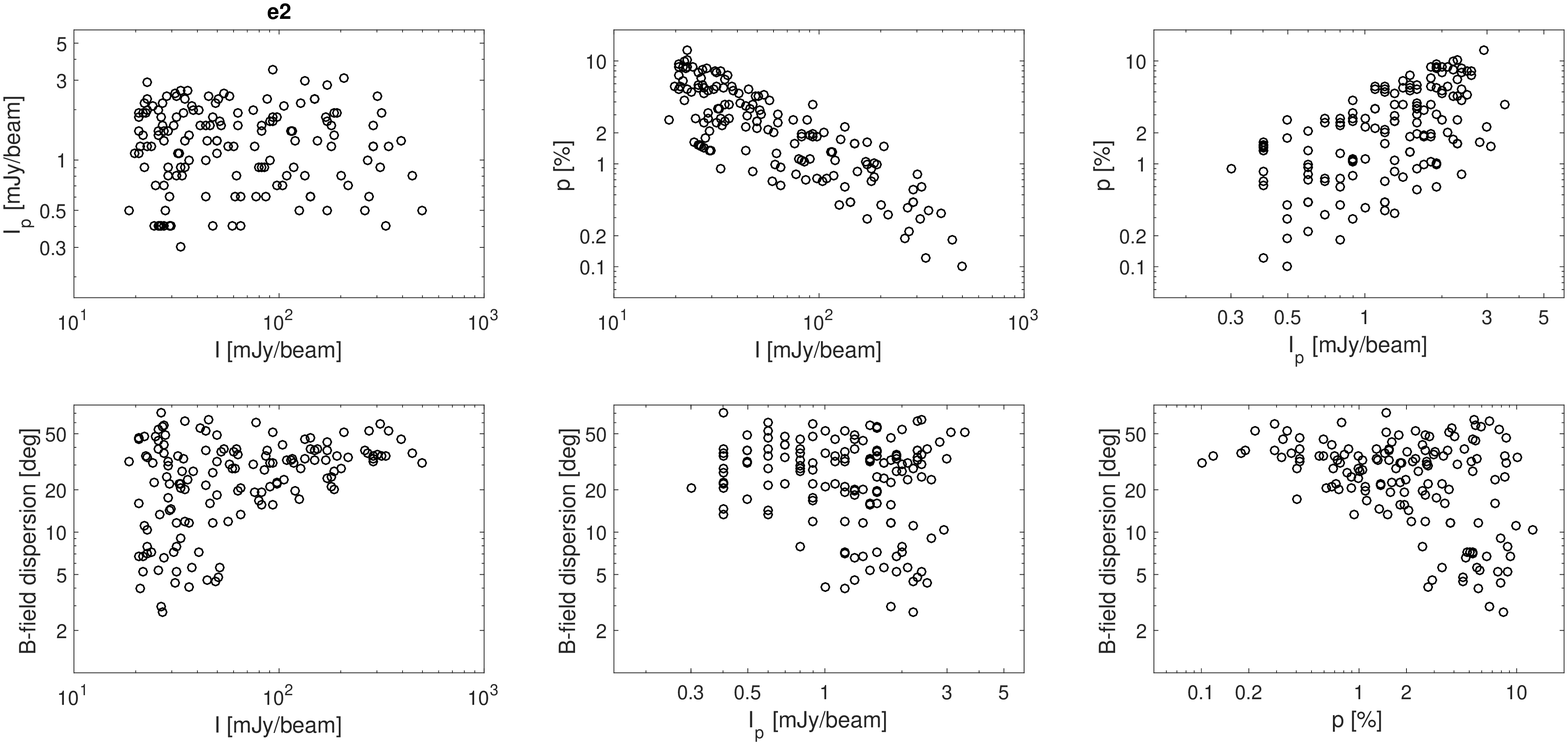}
\includegraphics[width=18cm]{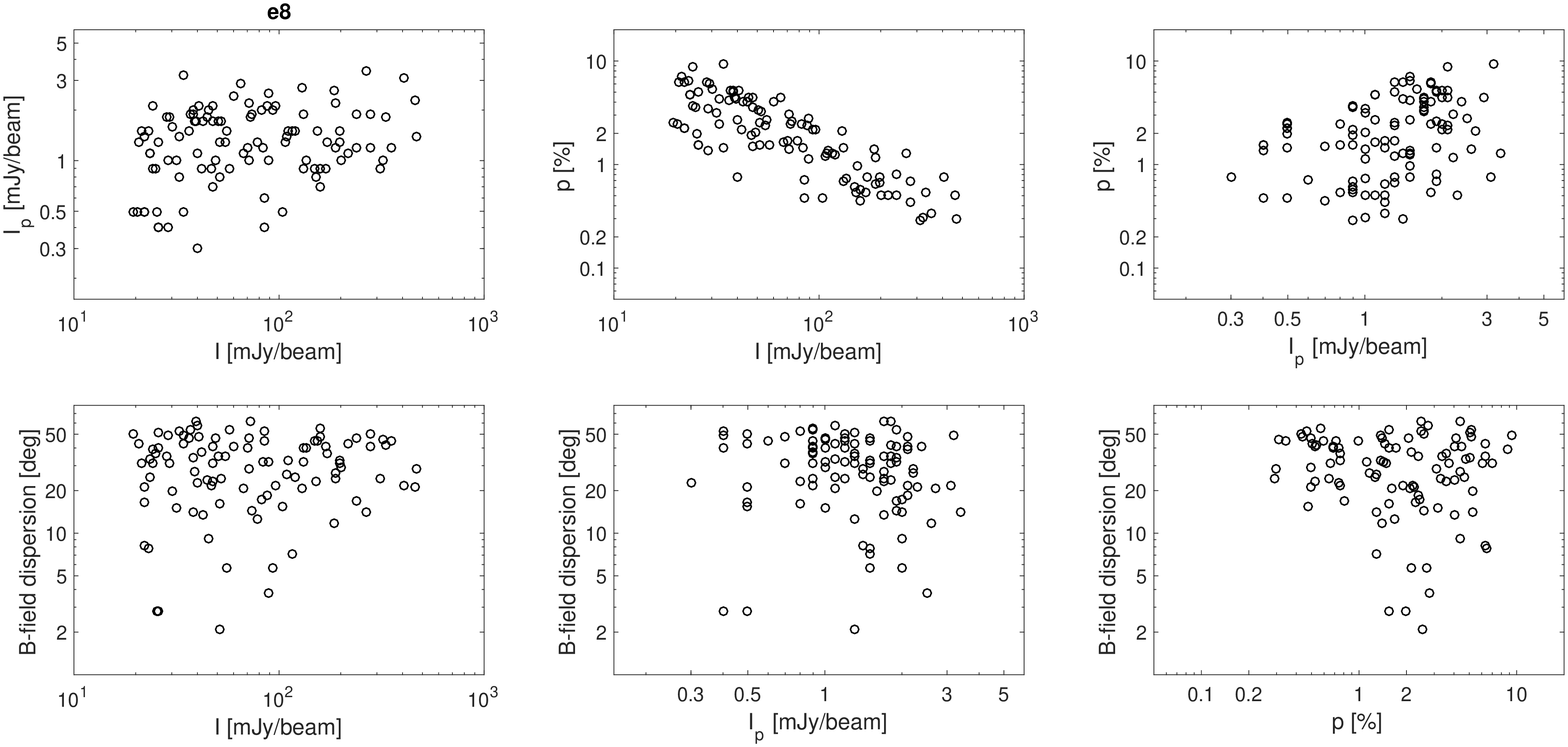}
\includegraphics[width=18cm]{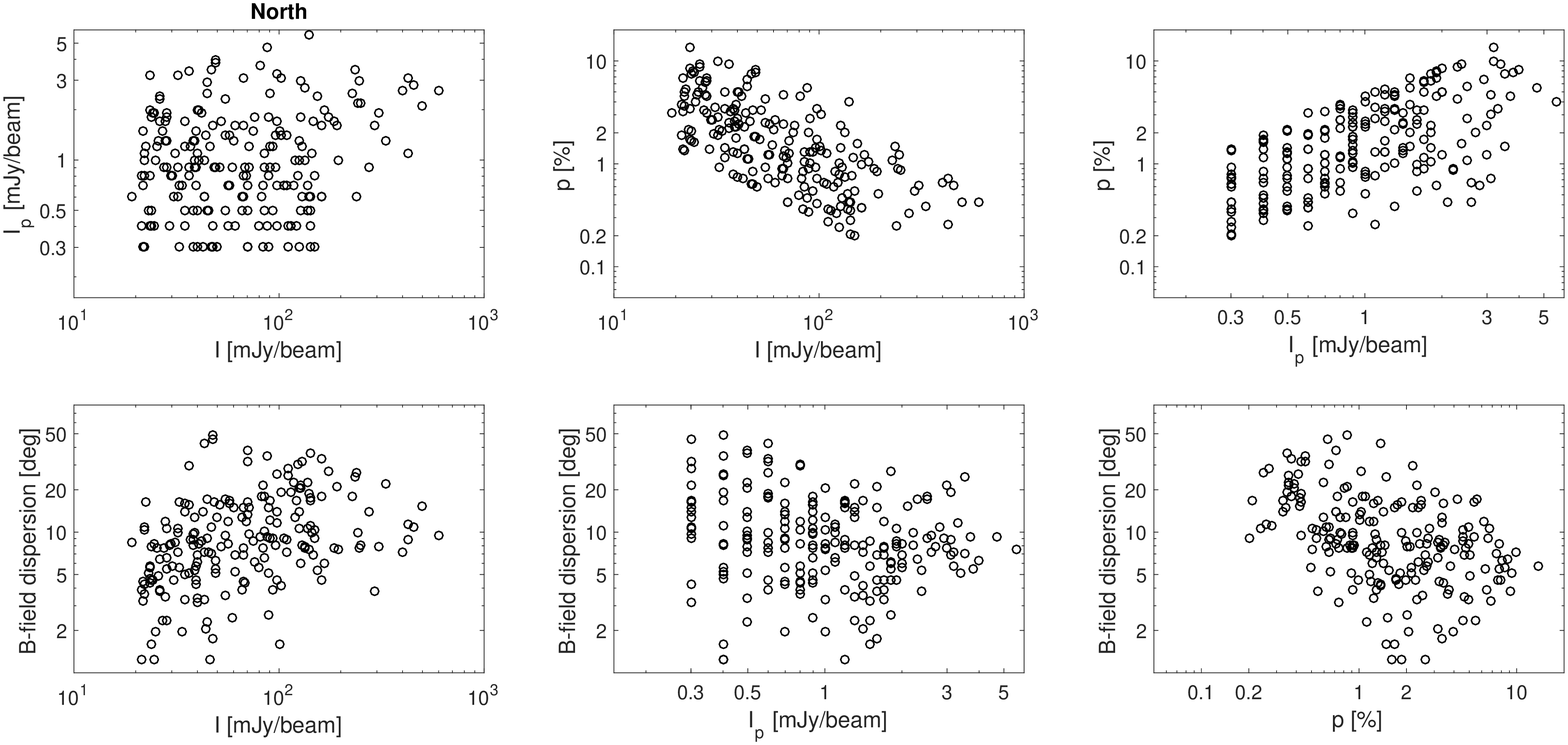}
\caption{\scriptsize All possible correlations among Stokes $I$, absolute polarization $I_p$, polarization percentage $p=I_p/I$, and local B-field dispersion $\mathcal{S}$ for W51 e2 (top two rows), e8 (middle two rows), and North (bottom two rows)
observed with ALMA.}
\label{figure_all_correlations_polarizations} 
\end{figure}

%
%

%
%
%
{\it Facilities:} \facility{ALMA, SMA}.

\acknowledgments{
The authors thank the referee for valuable comments and suggestions that further improved this manuscript.
This paper makes use of the following ALMA data: \\
ADS/JAO.ALMA\#2013.1.00994.S. ALMA is a partnership of ESO (representing its member states), NSF (USA) and NINS (Japan), together with NRC (Canada), MoST and ASIAA (Taiwan), and KASI (Republic of Korea), in cooperation with the Republic of Chile. The Joint ALMA Observatory is operated by ESO, AUI/NRAO and NAOJ.
PTPH acknowledges support from the Ministry of Science and Technology (MoST) in Taiwan through grant MoST 
105-2112-M-001-025-MY3.
Y-WT is supported through grant MoST 103-2119-M-001-010-MY2. PMK acknowledges support from
MoST 103-2119-M-001-009 and from an Academia Sinica Career Development Award.

}
%
%

\bibliographystyle{apj}                       
\bibliography{w51}
\end{document}